\documentclass[a4paper,fleqn,usenatbib]{mnras}

\usepackage{newtxtext,newtxmath}

\usepackage[T1]{fontenc}
\usepackage{ae,aecompl}

\newcommand{\bDiamond}{\mathbin{\Diamond}}
\newcommand{\bLozenge}{\mathbin{\blacklozenge}}

\usepackage{graphicx}	
\usepackage{amsmath}	
\usepackage{amssymb}	
\usepackage{multirow}

\title[Clarifying special Guest Stars]{A Search for the modern counterparts of the Far Eastern guest stars 369 CE, 386 CE, and 393 CE}

\author[Hoffmann et al.]{
Hoffmann, Susanne M.,$^{1}$\thanks{E-mail: susanne.hoffmann@uni-jena.de (PAF, FSU)}
Vogt, Nikolaus,$^{2}$
\\
$^{1}$Physikalisch-Astronomische Fakult\"at, Friedrich-Schiller-Universit\"at Jena, Germany\\
$^{2}$Instituto de Física y Astronomía, Universidad de Valparaíso, Chile\\
}

\date{Accepted 2020 June 30. Received 2020 June 15; in original form 2020 May 20}

\pubyear{2020}

\begin{document}
\label{firstpage}
\pagerange{\pageref{firstpage}--\pageref{lastpage}}
\maketitle

\begin{abstract}
 In this study, we apply our previously developed method to investigate ancient transient sightings in order to derive consequences for modern astrophysical problems. We present case studies of three observations of so called `guest stars' in the 4th century CE which lasted several months each. These three observations had been suggested and discussed as possible supernovae but slow novae are also viable alternatives. Our careful re-interpretation of the historical texts and the currently known objects in the given fields shed new light on this topic. In particular, for the two events in 386 and 393 CE we suggest possible supernova identifications, while in all three cases there are interesting candidates for past classical or recurrent nova eruptions among known cataclysmic variables (CVs) and/or symbiotic stars. The transient of 369, we suggest to explain as classical and possibly recurrent nova instead of a supernova. The most plausible candidates are BZ~Cam, a CV with a possible nova shell, or CQ~Dra, a naked eye multiple system perhaps able to permit an overwhelmingly bright nova with day-time visibility. 
\end{abstract}

\begin{keywords}
(stars:) binaries (including multiple): close -- (stars:) novae, cataclysmic variables -- history and philosophy of astronomy 
\end{keywords}



\section{Introduction}
Previously, we developed a method to evaluate historical records of transients in order to estimate their nature and modern counterpart. We tested this method on 24 historical events, identified two corrupt entries and found a highly likely new supernova in this data as well as several suggestions of cataclysmic variables (CVs). The steps of evaluation are described in our earlier Papers\,3 to 5 after two preliminary studies: 
 \begin{enumerate} 
	\item a review on published attempts of nova identifications among Far Eastern guest stars \citep{vogt2019},
	\item a careful analysis of the positions given in the historical text as sketched for the known cases in \citep{hoffmann2019}, 
	\item the definition of search areas and look for CVs therein; \citet{hoffmannVogtProtte} performed this successfully for a test sample of 26 ancient events.
	\item a careful individual considering of the CV findings in these fields; \citet{hovoMNRAS2020} performed this for the test sample selected in \citep{hoffmannVogtProtte}, 
	\item a search for alternative counterparts in these fields such as planetary nebulae, symbiotic stars, supernova remnants and pulsars; performed in \citet{hoffmannVogtLux} for the above mentioned test sample. 
 \end{enumerate}

\begin{figure}
    \caption{Visual display of the number of objects considered for this study (logarithmic scale). The two numbers in the first three bars display the amount of objects of each type in our search fields, e.\,g. of all 11,223 CVs enrolled in the VSX only 1,841 were in our search fields. The totals of SNRs and PSRs in Simbad is a little bit higher than displayed here because there are some few (order of ten) object names without coordinates which we did not consider.}
    \label{fig:stat}
	\includegraphics[width=\columnwidth]{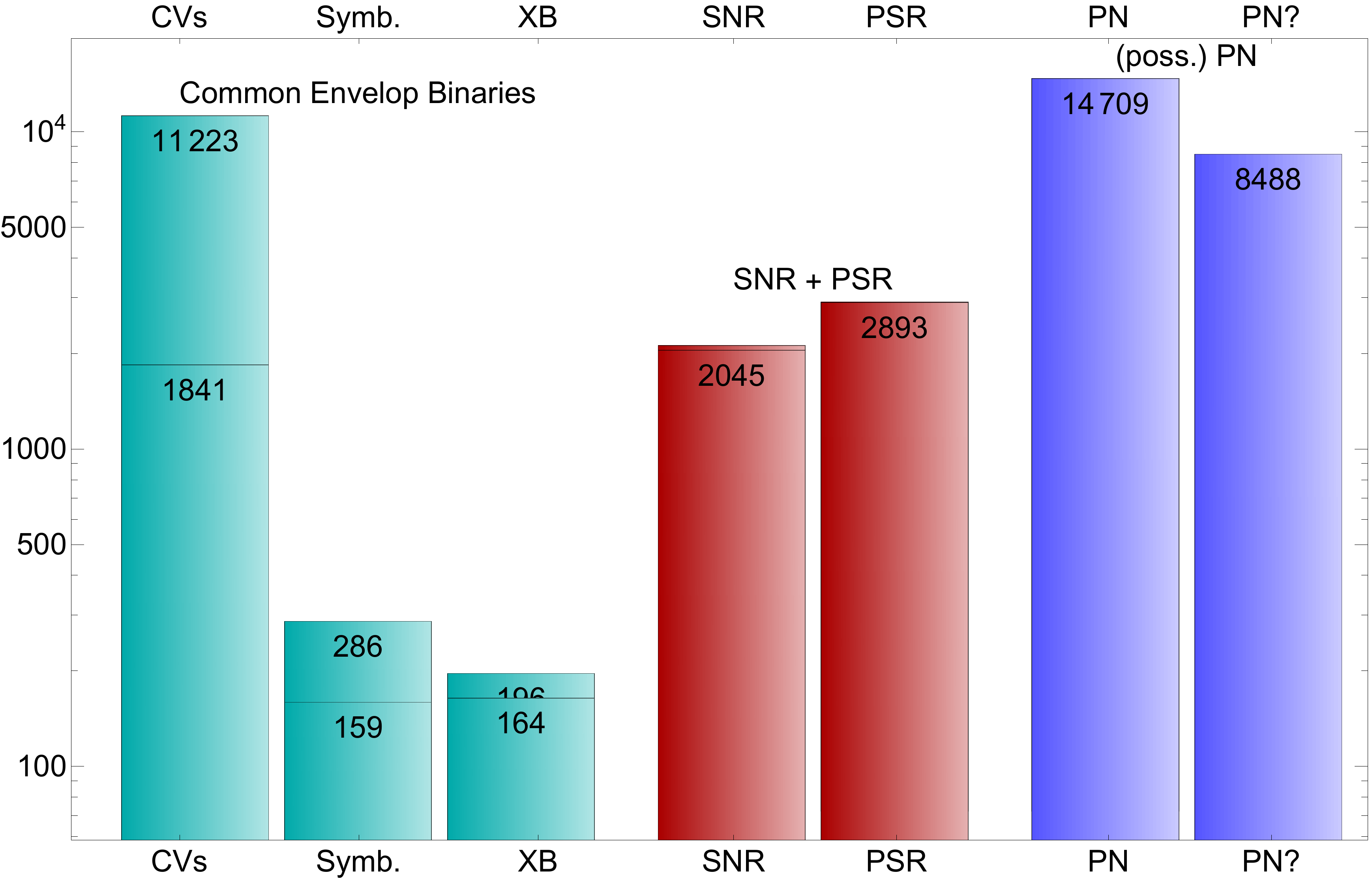}
\end{figure}

 This way, we aim to identify modern counterparts of possible stellar transients among Far Eastern guest stars (no matter whether they are called as such or referred to as `new star' or simply `star'). Our efforts to establish a special sort of (historical) `big data' analysis: A first selection was followed by a careful handpicking of the found objects of different type where the analysis includes (almost) all known type of brightening variability (except microlensing which shall be considered in the future). The totals of object types we already described in our previous studies \citep{hovoMNRAS2020,hoffmannVogtLux} are here only briefly visualised in Fig.~\ref{fig:stat}. As these computations of some ten thousand objects are already prepared, we are now starting to use this dataset to study the modern counterpart of the other $\sim160$ events in our list of potential stellar transients that had not been part of the test sample. From this set of $\sim160$ events, we now study three appearances of extraordinarily long duration in the 4th century.  

 \textbf{Goal:} There are three sightings, reported for several months each, in the years 369, 386, and 393~CE. The wording of the records in three modern translations is displayed in Tab.~\ref{text:369}, \ref{text:386} and \ref{text:393}, and we have added the sources as written by the modern translators and their comments on suggested identifications (if given). Hsi and Ho use an older transliteration style than Xu et al. but apart from this, all three modern translations agree.

 \subsection{State of the art} 
 Concerning the astrophysical research on these objects, the leading author is F. R. Stephenson and colleagues: In their book `The Historical Supernovae' \citet{steph77} resume on page~65: `In looking for possible supernovae, we are thus left with the stars of AD\dots 369, 386, 393 \dots\ .' On pages~112 and 113, after a careful analysis of the SNRs in the respective areas of the latter two sightings, they resume: `Although there is ample vindication for the essential reliability of the astronomical records of the Chin Dynasty, the brevity of the descriptions of the two `long' duration guest stars of AD~386 and 393 makes an interpretation of their exact nature and behaviour almost impossible.' However, Stephenson continues his efforts to identify the remnants of these sightings in his more recent book `Historical Supernovae and their Remnants' \citep[chap.\,10.3]{stephGreen}. On page~179 they summarise a very brief mentioning of event~369: `However, because of the considerable uncertainty in the position of the guest star, a search for a remnant would be profitless.' For the event~386, they suggest to study SNR G11.2-0.3 and `Further observations of both G11.2-0.3 (\dots) and its central pulsar will help confirm or refute the proposed association with the possible SN of AD~386.' On pages~184 and 186 they state the difficulty to identify a certain SNR for the appearance in 393 because this area is close to the galactic centre and rich of objects of any type, even supernova remnants. They refer to their earlier work \citep{steph77} which knew already 7 SNRs within the described asterism while the research of a quarter of century between this and \citet{stephGreen} brought up new knowledge and some exchange but the `current catalogue lists 14 SNRs in the same region'. That is why, their conclusion is that `it is not possible to identify a clear candidate for the remnant of the SN of AD~393.'

\begin{table}
 \caption{Three variants of translation for the record of 369. The second row copies the sources of the author(s) and the third row his/their comment. `Zigong' is the modern transliteration of `Tzu-wei'.} 
 \label{text:369}
 \begin{tabular} {p{.28\columnwidth}p{.28\columnwidth}p{.28\columnwidth}}  
 \citep{xu2000} & \citep{ho} &\citep{hsi} \\ \hline 
 Emperor Haixigong of Jin, 4th year of the Taihe reign period, 2nd month. \\
 A guest star appeared at the west wall of Zigong through the 7th month (Aug 19 to Sep 17), when it was extinguished. 
  & 
   \dots\ a guest star appeared at the western wall of the Tzu-Wei (Enclosure). It went out of sight during the seventh month [19th August to 17th September].
  & \dots\ a guest star was seen in the western constellation Tzu-wei and it disappeared by the 7th month.
  \\ \hline
   Jin shu, Tianwen zhi, ch.\,13; Song shu, Tianwen zhi, ch.\,24
  & CS 13/20b; SShu 24/27a; WHTK 294/9b; B(1) W132; L; Hsi.
  & Chin-shu, T'ung-shih and T'ung-k'ao 
 \\\hline
  -- 
 & This was once thought to correspond with radio source ($\alpha=23^\textrm{h} 21^\textrm{m}$,$\delta=58\degr$), but Hsi disagrees.
 & Shklovskii and Parenago have noted that a radio point source is observed at this position. Possibly a supernova.
 \\ 
 \end{tabular} 
\end{table} 
\begin{table}
 \caption{Three variants of translation for the record of 386. The second row copies the sources of the author(s) and the third row his/their comment. Remark: Hsi's booklet with the `Catalog of Ancient Novae' has the imprinted time stamp 1958 but in arXiv it is referred to as 1957. It is always the same list which is referred to.} 
 \label{text:386}
 \begin{tabular} {p{.28\columnwidth}p{.28\columnwidth}p{.28\columnwidth}}  
 \citep{xu2000} & \citep{ho} &\citep{hsi} \\ \hline 
 Emperor Xiaowu of Jin, 11th year of the Taiyuan reign period, 3rd month. \\
 There was a guest star in Nandou [LM8] that lasted until the 6th month (Jul 13 to Aug 10), when it disappeared. 
  & \dots\ a `guest star' appeared at the Nan-Tou (eighth lunar mansion). It went out of sight during the sixth month [13th July to 10th August].
  & \dots\ a guest star was in the constellation of the Southern Dipper. It faded away in the 7th month.
  \\ \hline
   Jin shu, Tianwen zhi, ch.\,13; Song shu, Tianwen zhi, ch.\,25
  & CS 31/21a; SShu 25/5a; WHTK 294/9b; B(1) W134; L; Hsi.
  & Chin-shu, T'ung-shih and T'ung-k'ao\\\hline
  -- 
 & It is said to be found near NGC 6644; cf. Hsi Tsê-Tsung (1958) 
 & -- 
 \end{tabular} 
\end{table} 
\begin{table}
 \caption{Three variants of translation for the record of 393. The second row copies the sources of the author(s) and the third row his/their comment.} 
 \label{text:393}
 \begin{tabular} {p{.28\columnwidth}p{.28\columnwidth}p{.28\columnwidth}}  
 \citep{xu2000} & \citep{ho} &\citep{hsi} \\ \hline 
 Emperor Xiaowu of Jin, 18th year of the Taiyuan reign period, 2nd month. \\
 There was a guest star in the middle of Wei [LM6] that lasted until the 9th month (Oct 22 to Nov 19), when it was extinguished.
  & \dots\ a `guest star' was seen within the Wei (sixth lunar mansion). It went out of sight during the ninth month [22nd October to 19th November].
  & \dots\ a guest star was found in the constellation Wei (Queue) and disappeared by the 9th month. 
  \\   \hline
    Jin shu, Tianwen zhi, ch.\,13; Song shu, Tianwen zhi, ch.\,25
  & CS 12/21a; SShu 25/7a; WHTK 294/9b; B(1) W136; L; Hsi.
  & Chin-shu, T'ung-shih and T'ung-k'ao\\\hline
  -- 
 & regarded as the nova NGCII 4637; cf. Hsi Tsê-Tsung (1958)
 & studied by Williams and Biot, but their results differ. Ours is in agreement with Biot's. This nova is found in the vicinity of NGC II 4637 and a Wolf-Rayet star, $-40\degr10$ 919.
 \\ 
 \end{tabular} 
\end{table} 

 \subsection{Open questions} Fig.\,2 of \citet{hovoMNRAS2020} shows that the typical decline of novae observed in modern time varies from several days to several months (based on data by \citet{strope2010}). In this figure, we plotted the $t_3$ time, i.\,e. the time a nova needs to fade from peak by 3~magnitudes. Most of the novae need less than 40~days and 25~days are a rather typical time but there are also cases with 250~days (8~months) and more. That is why, the \textit{longue-durée} guest stars reported in the 4th centuries which are already suggested and often discussed as supernovae could equally likely be slowly declining novae.  

 Therefore, in our \citet{hoffmannVogtProtte}, we first included the event~369 (which does not yet have an SNR counterpart suggestion) in our list of events to analyse with higher priority. Nevertheless, in a second step, we suggested to postpone this sighting to a later study for the same reason as given by Stephenson: The field which must be included in our search should be enormously huge. Additionally, as it has been suggested as supernova and not nova, it was initially unclear if our efforts will yield a success when we were seeking for CVs only. Being now trained with `historical big data analysis' in order to identify historical guest stars, we get back to this event and apply the same procedure also to the other two `long duration guest stars' (as Stephenson puts it) of the 4th century. We do not expect to finally solve the whole puzzle what the ancient astronomers saw in these cases but we hope to provide a realistic tendency.  
 
 \subsection{Types of novae} 
Novae are grouped in the categories Na, Nb, and Nc for fast, moderately fast, and slow decline \citep[Tab.\,1.1 on p.\,3]{bode}, \citep[Tab.\,5.4, p.\,263]{warner1995}. Due to their long visibility of several months, only slow novae are valid candidates for the here considered historical records. As fast nova amplitudes exceed slow nova amplitudes by a few magnitudes \citep[Fig.\,2]{dellaValle1991}, \citep[Fig.\,5.4, p.\,265]{warner1995}, our magnitude limit of $18~mag$ for the modern counterpart defined in \citet{hoffmannVogtProtte} for the general case does not apply in these special cases: Of course, the counterpart CV should be bright enough to reach naked eye visibility with a typical outburst amplitude. However, it is not sufficient to be hardly visible (5~mag) but there should be the possibility to decline by (at least) 2~mag in a couple of months. This puts the magnitude limit $m_{CV}$ to $m_\textrm{peak}+A+2$~mag with typical amplitudes of slow novae of $A\leq10$~mag \citep[Chap.\,5.2]{warner1995}. Therefore, our cataclysmic variable (CV) candidates for Nc-type should be brighter than 14~mag.
 
 \textbf{Symbiotic stars} are common envelope binaries accreting from a red giant but not necessarily onto a white dwarf \citep{munari2019}, they are called Z~And-type. Typically they show variabilities by less than 1 or up to 3~mag. Only some of the systems of this very inhomogeneous type can permit nova eruptions, e.\,g. KT~Eri which is known as Na-type classical nova \citep{hoffmannVogtLux}. As for cataclysmic binaries, Nc-type novae from symbiotic binaries decline much slower and have smaller amplitudes: The maximum amplitude is $\Delta V=11.0$~mag (from 19.5 flared up to 8.5~mag) \citep{goranskij2010} with a $t_3\sim220$~days and typical amplitudes are 7 to 10~mag \citep{hoffmannVogtLux}. It had been a common belief that nova ejections from symbiotic stars are considerably decelerated by the common envelope of the binary but the revision of this paradigm by \citet{munaribanerjee2018} in a case study of Nova Sco~2014 has proven this not the general case. They mention that further cases `hidden in past novae' may improve our understanding.

 Not to be confused with these classical novae from symbiotic stars is the very rare case of \textbf{symbiotic novae} (SyN) which is characterized by a long plateau phase of the outburst with decline times of several years as in case of V1016~Cyg or V4368~Sgr \citep[p.\,86]{munari2019} and which could last even a century as in case of BF~Cyg. There are only $\sim10$ symbiotic novae currently known which outbursted in historical time. As we are looking for objects able to explain a sighting of only several months, symbiotic novae of Z~And-type stars are not considered as candidates in our search. 

  Classical novae from symbiotic binaries, instead, are highly interesting candidates. Most famous classical novae from symbiotic binaries are RS~Oph, T~CrB, V745, and V3890~Sgr \citep[p.\,89]{munari2019} which are all known to have had more than one outburst and are, therefore, recurrent novae. Because of the smaller amplitudes of novae from symbiotic systems, Z~And-type counterparts for naked eye sightings by historical astronomers should in general also be 14~mag normal level brightness \citep{hoffmannVogtLux}. 
 
 A special outcome of our studies could be new cases of the novae of Nr-type, the \textbf{recurrent novae} \citep{darnley2019}. Nr-type novae split in three subclasses \citep[chap.\,5.9]{darnley2019,warner1995}: ($i$) the T~Pyx- or RS~Oph-type with M giant secondaries ($P_\textrm{orb}\sim 400$~d), ($ii$) U~Sco class with slightly evolved late type main sequence secondaries ($P_\textrm{orb}\sim 2$~d), and ($iii$) T~CrB-type with M dwarf secondaries ($P_\textrm{orb}\sim0.1$~d). Recurrent novae with evolved secondaries (of T~CrB- and U~Sco-type) lie 2--5~mag below the normal amplitude $A$-log$t_2$ relationship of classical novae. Nr eruptions do not peak to always the same magnitude but the peak brightness can vary by 1 or 2 magnitudes \citep{mayall1967,schaefer2010}. Most of currently known Nr in the Milky Way, M31 and the LMC have been found to erupt in intervals of 4 to 38~years \citep[Fig.\,2]{darnley2019}. According to the same author (his Table~1), there are 3 cases among 10 in the Milky Way with recurrent periods between 80 and 98 years. Apparently, there is no upper limit of the recurrence periods of Nr. Thus, we expect that increasing the temporal baseline of observations will lead to a broader distribution of possible recurrence periods, possibly even centuries or millennia if we manage to suggest proper and certain identifications of guest stars. 

\section{Applying our method}  
As shown in our earlier papers, the approach to find out what have caused the `guest star' starts with carefully re-interpreting the historical text. That is why, we cite the exact wordings in Tab.~\ref{text:369}, \ref{text:386}, and \ref{text:393} and the analysis of them is performed in the following two subsections. Subsequently, the next steps are:
\begin{enumerate} 
 \item Defining search areas and partition them in search circles for VSX input: Tab.~\ref{tab:fields}.
 \item Probing CVs, X-ray binaries, symbiotic stars, and nebulae: Figures~\ref{fig:all369} to \ref{fig:all393} in the Appendix, and
 \item Revisiting SNRs and PSRs (cf. Appendix). 
\end{enumerate} 
 
 The Figs.~\ref{fig:369}, \ref{fig:all386}, and \ref{fig:393} present the areas in the sky, and mark possible counterparts as explained in the following two subsections per event. Table~\ref{tab:allCVs} lists the selected counterpart candidates. 
 
 \textbf{Elaborated method:} We should stress that we give the search circles in Tab.~\ref{tab:fields} (and in the maps) just for the consistence with our earlier suggestions \citep{hoffmannVogtProtte} to query the Variable Star Index (VSX) of the AAVSO and for easy reproducibility by other scholars. In this study, we did not use them directly. Instead, we used a more sophisticated way to ($i$) download the current state of the lists of objects included in the VSX and CDS Simbad, ($ii$) plotted them into our maps (cf. Appendix) and ($iii$) derived the object lists in our Tab.~\ref{tab:allCVs} manually by only regarding the maps. This way allows to immediately regarding the position of the object with respect to the asterisms lines and, thus, avoids to do this in a further step of filtering (method already described in \citet{hoffmannVogtLux}). The circles are good for a first quick search but our studies use a more extensive method with interactive star charts. 

 \subsection{Position of event 369}
 The position in the first record is the most difficult one: The appearance is said to have been located in the asterism of Zigong, the Palace of Purple, which extends to almost the complete circumpolar area (Fig.\,8 in \citet{hoffmann2019}). As already suggested by \citet[p.\,81]{steph77}, the search for counterparts close to the border of the celestial enclosure should be performed in a stripe around declination $+65\degr$ but in contrast to them, we reduce the right ascension to the half circle: Fortunately, it is given that the guest star was seen at the \textit{western} wall which reduces the search field. Nevertheless, it has to be clarified what `west' in this case means: From the pole any direction is `southwards' and as the ever-visible circle turns continuously this description cannot be meant with regard to the horizon. The answer has to be found in the systematic of the alignment of Heaven and Earth in Chinese belief and its projection into asterisms.  
 
  The concept of a `left' and `right' or `east' and `west' division of the sky is very old in China. In tombs, temples and other archaeological proofs it can be traced back to neolithic times, \citep[from Fig.\,2.1 on page\,39 to Fig.\,3.2 on pages 84-87]{pankenier2013}. \citet[Fig.~3.2]{pankenier2013} shows the ground view of Chinese temples and the Forbidden City in the 2nd millennium which helps to understand the asterism of Zigong: its main gate should be considered the southern end of the enclosure and the guest houses are in the north. Still, it remains the question for `east' and `west' whether the `map' of the celestial projection of the palace is considered as map of the sky or map of the earth, in other words: Is `west' left or right? To answer this question, we should consider other asterisms: ($i$) Next to Arcturus, the single star asterism of the Great Horn, there are the asterisms of the Left and the Right Assistant Conductors out of which the right one is the western one. The same principle of naming can be found in the asterisms Yingshi [Align-the-Hall] and Dongbi [Eastern Wall] in Pegasus \citep[p.\,129-137]{pankenier2013}. ($ii$) There is a very ancient division of the sky into three dominions, namely of the asterisms of the Dipper (UMa) as north, the Fire Star (Antares in Sco) as east, and the Triaster (Ori) as the west which is preserved in several engravings, cf. \citet[p.\,57--58]{pankenier2013}. This refers to the grouping of lunar mansions as shown in the map of our \citet[Fig.\,5]{hoffmannVogtProtte}: ($iii$) Within the Chinese concept of the Four Holy Beast in the sky, the Azure Dragon of the East contains Antares as his heart and the White Tiger of the West contains Orion. Thus, the West Wall of the Zigong Enclosure (equaling the right wall) always has been considered the half circle in the constellations Draco, Ursa Major, and Camelopardalis as it is defined today and applied by \citet[p.\,179]{stephGreen}.  

 As it is difficult to partition such a long polygon (a stripe) into many search circles, we dropped this step of our method here. Instead, we map all object lists of possible counterparts into the celestial map of the half circumpolar area and display the asterism line of the `west wall of Zigong' along which the guest star was seen. A look at this map immediately leads to a list of objects which would most likely be described by human observers as `at the west wall' and not at a nearby neighbouring asterism. 
 
 \begin{table*}
 \caption{Search fields for the quest stars to find modern counterparts of the ancient transients. The areas in the last column are the areas covered by the circles and, thus, overestimated: The circle in 393 covers much more than `the middle of Wei' and the half ring for 369 covers much more than the area around the line which is called `west wall of Zigong'. In both cases the search is further constrained by regarding the maps (Fig.~\ref{fig:369} and \ref{fig:393}).} 
 \label{tab:fields}
  \begin{tabular}{ccllccc cc}
   year &    & duration & asterism  & HIP & RA2000 & DE2000 & radius/ \degr & area/ $\degr^2$\\
    \hline
   369  &    & 5 \text{months} & \text{west wall of Zigong}  & &0 &90 & 32 & 1614\\
   &\multicolumn{7}{p{.61\textwidth}}{from this circle we subtract the inner circle of 13\degr\ and 
 consider only the half from RA 1\,h to 15\,h~25\,m; search field already defined in \citet{hoffmannVogtProtte}.}\\
  \hline
   386  & 1. & 3 \text{months} & \text{in Nandou} & 92927 & 283.995 & $-28.1302$ & 4 & \multirow{4}{5em}{101} \\
   386  & 2. & 3 \text{months} & \text{in Nandou}   & 90496 & 276.993 & $-25.4217$ & 3 & \\
   386  & 3. & 3 \text{months} & \text{in Nandou}   & 92041 & 281.414 & $-26.9908$ & 3 & \\
   386  & 4. & 3 \text{months} & \text{in Nandou}   & 89341 & 273.441 & $-21.0588$ & 3 & \\
   \hline
   393  &    & 7 \text{months} & \text{in the middle of Wei [Sco]} & 84638 & 259.55 & $-39.6908$ & 5 & 78\\
   \end{tabular}
 \end{table*} 

 \subsection{Positions of event 386 and event 393} 
 These two events leave no doubt how to proceed; we can strictly follow the scheme as defined in \citet{hoffmannVogtProtte}: The event in 386 is reported in the asterism of Nandou, the Southern Dipper. It is part of the IAU-constellation of Sagittarius. As the text does not allow to narrow the position, we have to search within the whole asterism area. Thus, our search circles for the VSX-search are defined to cover the whole asterism of Nandou. Any type of area (be it Stephenson's polygons or our VSX-circles) will have to be defined with a little extend towards the neighboring asterism, e.\,g. until the half way between. However, looking at the sky and discovering a `new star' an astronomer would likely describe an appearance between two asterism lines as `between X and Y' or `in space of X and Y' rather than `in X'. Therefore, after defining our search circles and looking for possible counterpart objects, we will plot these objects, asterisms, and search circles into star charts and evaluate the likelihood of an object matching the preserved description. Yet, `Nandou' is not only an asterism but also a lunar mansion. Thus, this likelihood or matching margin is bigger to the north and south than to the east and west (towards the neighbouring lunar mansions). However, in this case, probably the asterism is meant and not the lunar mansion because otherwise it would have likely been written the determining term `xiù' in connection with the name.

 The event in 393 appeared `in the middle of Wei' where `Wei', the Tail (of the Chinese Dragon), designates the region which is considered the tail of the IAU-Scorpius. As this asterism is U-shaped, we define our search circle centered in the middle of the `U' and extending more or less to the asterism line. It does not matter how perfectly the lines of the asterism and the circle match because the text reports `in the middle' and a human observer would not describe something `in the middle' if it was at the edge of a circle. Therefore, the best way to proceed in our search will be to define a circle bigger than needed, plot it into a map and, then, select our possible targets at a glance, i.\,e. having the same perception as the ancient astronomer. 
 
 \begin{table*}
 \caption{CVs, X-ray binaries, and symbiotic stars as possible counterparts of the sightings reported in the 4th century. The ID numbers in the second column are displayed in the maps Fig.~\ref{fig:369} to \ref{fig:393}.} 
 \label{tab:allCVs}
 \scriptsize
 \begin{tabular}{crlllcp{16ex}ccp{36ex}}
 year & ID& star name & RA2000$/\degr$ & DE2000$/\degr$ & & type & period/ d & magnitude(s) & commentary \\
 \hline
 369 & \multicolumn{9}{p{.93\textwidth}}{\footnotesize Listed are only those stars directly at the asterism line and not all in the search field as displayed in Fig.~\ref{fig:all369}. BZ~Cam is not directly at the asterism line but included because it is the only object with a nearby nebula, cf. Fig.~\ref{fig:all369}. The objects 1 to 5 are highlighted in Fig.~\ref{fig:369}}\\
 369 & 1. & \text{LU Cam} & 89.5748 & 67.8962 & \text{Cam} & \text{UG} & 0.149969 & \text{14 - $<$16.0 V} & \text{possible} \\
 369 & 2. & \text{1RXS J041924.8+653006} & 64.8482 & 65.5012 & \text{Cam} & \text{UG} & \text{--} & \text{13.9 - 16.5 CR} & \text{} \\
 369 & 3. & \text{ASAS J071404+7004.3} & 108.519 & 70.0718 & \text{Cam} & \text{CV:} & 22.44 & \text{11.2 - 13.3 V} & \text{} \\
 369 & 4. & \text{BZ Cam} & 97.392 & 71.0769 & \text{Cam} & \text{NL/VY} & 0.15353 & \text{12.5 - 14.1 B} & \text{highly likely nova, embedded in multiple shell-nebula} \\
 369 & 5. & \text{CQ Dra} & 187.528 & 69.2011 & \text{Dra} & \text{ZAND} & \text{1703 d(4.66 y)} & \text{4.90 - 5.12 V} & \text{highly likely} \\
 369 & 6. & \text{SDSS J100516.61+694136.5} & 151.319 & 69.6935 & \text{UMa} & \text{DQ:} & \text{--} & \text{17.9 - 21.0 r} & \text{too faint} \\
\hline
 \multicolumn{10}{p{\textwidth}}{\footnotesize Listed are all bright common envelope and cataclysmic binaries returned by our queries in the VSX and Simbad.}\\
 386 & \multicolumn{9}{p{\textwidth}}{\footnotesize Fig.~\ref{fig:all386} displays all objects, Fig.~\ref{fig:386} this selection.}\\
 386 & 1. & \text{V5850 Sgr} & 275.747 & -19.2366 & \text{Sgr} & \text{NA:} & \text{--} & \text{11.3 - 21.4: CR} & \text{too faint} \\
 386 & 2. & \text{V1151 Sgr} & 276.335 & -20.1962 & \text{Sgr} & \text{NA:} & \text{--} & \text{10.0 - 19.7: B p} & \text{too faint} \\
 386 & 3. & \text{ASASSN-15rc} & 284.025 & -26.3395 & \text{Sgr} & \text{UG} & \text{--} & \text{15.56 - $<$16.0 V} & \text{this case: too faint} \\
 386 & 4. & \text{ASASSN-17mj} & 286.591 & -28.8524 & \text{Sgr} & \text{UG} & \text{--} & \text{13.9 - 17.8 V} & \text{this case: too faint} \\
 386 & 5. & \text{OGLE-BLG-DN-1057} & 279.103 & -23.9099 & \text{Sgr} & \text{UG} & \text{--} & \text{15.2 - 17.3 Ic} & \text{in M22, unlikely and in this case: too faint} \\
 386 & 6. & \text{OGLE-BLG-DN-1017} & 274.781 & -24.2814 & \text{Sgr} & \text{CV} & \text{--} & \text{16.1 - 16.4 Ic} & \text{likely too faint} \\
 386 & 7. & \text{OGLE-BLG-DN-1040} & 275.249 & -26.1251 & \text{Sgr} & \text{CV} & \text{--} & \text{17.5 - 18.0 Ic} & \text{likely too faint} \\
 386 & 8. & \text{V1223 Sgr} & 283.76 & -31.1638 & \text{Sgr} & \text{DQ} & 0.140244 & \text{12.3 - $<$16.8 V} & \text{} \\
 386 & 9. & \text{OGLE-BLG-DN-0985} & 274.091 & -23.345 & \text{Sgr} & \text{CV} & \text{--} & \text{17.3 - 18.1 Ic} & \text{too faint} \\
 386 & 10. & \text{MACHO 311.37055.2070} & 274.478 & -23.8405 & \text{Sgr} & \text{UG:} & \text{--} & \text{--} & \text{no data?} \\
 386 & 11. & \text{V4633 Sgr} & 275.419 & -27.527 & \text{Sgr} & \text{NA+E} & 0.125567 & \text{7.4 - 21: V} & \text{likely too faint} \\
 386 & 12. & \text{V0726 Sgr} & 274.89 & -26.8888 & \text{Sgr} & \text{NA+ELL} & 0.822812 & \text{10.5 - 19.4 V} & \text{too faint} \\
 386 & 13. & \text{IGR J18173-2509} & 274.342 & -25.1451 & \text{Sgr} & \text{DQ} & \text{0.06382:} & \text{17.2 - ? R} & \text{this case: too faint} \\
 386 & 14. & \text{IGR J18245-2452} & 276.137 & -24.8666 & \text{Sgr} & \text{LMXB} & \text{--} & \text{11.7 - 13.0 CR} & too faint (normally $\geq20.6$~mag)  \\
 386 & 15. & \text{MACHO 311.37557.169} & 274.674 & -23.9392 & \text{Sgr} & \text{AM:} & \text{--} & \text{15.3 - ? V} & \text{likely too faint} \\
 386 & 16. & \text{V5569 Sgr} & 282.515 & -26.4043 & \text{Sgr} & EA+BE+ ZAND  & 515. & \text{9.8 $--$ 12.1 V} & EM* AS 325; Be star, not nova  \\
 386 & 17. & \text{OGLE-BLG-ECL-000206} & 275.16 & -24.2418 & \text{Sgr} & \text{NL+E} & 0.161329 & \text{18.1 - 19.9 Ic} & \text{too faint} \\
 386 & 18. & \text{MACHO 172.31712.790} & 277.602 & -26.6322 & \text{Sgr} & \text{UG} & \text{--} & \text{16.6 - 19.5 V} & \text{this case: too faint} \\
 386 & 19. & \text{V3890 Sgr} & 277.68 & -24.0191 & \text{Sgr} & \text{NR+E} & 103.14 & \text{7.1 - 18.4 V} & \text{ } \\
 386 & 20. & \text{V0522 Sgr} & 282.002 & -25.374 & \text{Sgr} & \text{UGWZ:} & \text{--} & \text{12.9 p - 19.7: V} & \text{unlikely} \\
 386 & 21. & \text{Gaia18dkm} & 284.325 & -32.146 & \text{Sgr} & \text{UG} & \text{--} & \text{17.1 - 0.9 G} & \text{likely too faint} \\
 386 & 22. & \text{V5759 Sgr} & 271.391 & -20.3439 & \text{Sgr} & \text{ZAND+R} & 671. & \text{11.5 - 14.2 V} & \text{possible, unlikely} \\
 386 & 23. & \text{AS 327} & 283.319 & -24.383 & \text{Sgr} & \text{ZAND} & 823. & \text{12.6 - 13.5 V} & \text{possible, unlikely} \\
 386 & 24. & \text{V1988 Sgr} & 276.989 & -27.623 & \text{Sgr} & \text{ZAND:+SRB} & 94. & \text{12.0 - 14.0 V} & \text{possible} \\
 386 & 25. & \text{V3929 Sgr} & 275.245 & -26.8071 & \text{Sgr} & \text{ZAND} & \text{--} & \text{13.7 - $<$16.0 p} & \text{G 15.1 too faint, this year Gaia Alert} \\
 386 & 26. & \text{V2601 Sgr} & 279.509 & -22.6976 & \text{Sgr} & \text{ZAND} & 850. & \text{14.0 - 15.3 p} & \text{possible} \\
 386 & 27. & \text{V4641 Sgr} & 274.84 & -25.4072 & \text{Sgr} & HMXB/BHXB/ XN+ ELL+E & 2.81724 & \text{9.0 - 14.0 V} & unpredictable blackhole eruptions \\
 386 & 28. & \text{Hen 2-374} & 273.879 & -21.5897 & \text{Sgr} & \text{ZAND} & 820. & \text{14 - ? V} & \text{too faint} \\
\hline
 393 & \multicolumn{9}{p{\textwidth}}{\footnotesize Fig.~\ref{fig:all393} displays all objects, Fig.~\ref{fig:393} this selection.} \\
 393 & 1. & \text{V0643 Sco} & 261.481 & -40.9687 & \text{Sco} & \text{UGZ} & \text{--} & \text{13.0 - 14.2 V} & \text{possible} \\
 393 & 2. & \text{IGR J17195-4100} & 259.9 & -41.0149 & \text{Sco} & \text{DQ} & 0.1669 & \text{14.8 - 15.5 V} & \text{unrelated nebula} \\
 393 & 3. & \text{V0884 Sco} & 255.987 & -37.8441 & \text{Sco} & \text{ELL+HMXB} & 3.41161 & \text{6.51 - 6.60 V} & \text{embedded in huge nebula} \\
 393 & 4. & \text{V0902 Sco} & 261.535 & -39.0668 & \text{Sco} & \text{N:} & \text{--} & \text{11.0 - 17.0 p} & \text{too faint} \\
  \end{tabular} 
\end{table*} 

\section{Results and Discussion} 
  Concerning the duration given in the text (Tab.~\ref{tab:fields}), there is one specialty: The last of the three transients appeared in March 393 when the asterism of Wei culminated in the morning and the transient was traced until the sun reached the area of Scorpius and it, thus, disappeared in dusk at the end of September. This is, what is reported in the preserved text. Subsequently, there are two possibilities: In the end of December the asterism of Wei rises heliacally and becomes visible again in the morning sky. Either the guest star faded away and was really not visible any more, or further observations are lost because of astrological irrelevance (e.\,g. in December Jupiter also was in the lunar mansion of Wei, directly north of the given position in the asterism with the same name; this had not been the case in spring and summer of this year). Thus, the given duration of the event could be 7~months as given in the text or longer if it lasted unmentioned after heliacal rising. 
  
  In contrast, the transient in 369 is in the circumpolar area and, thus, always visible. There is no further information to cut down on the search field or the catalogues of objects. The event in 386 also does not have any magnitude mentioned. The appearance of the guest star is reported, when the asterism of Nandou is already visible since many months and when it vanishes, the asterism has its phase of longest visibility. Therefore, for the events 369 and 386 the given duration is much more reliable than for 393. 

 \subsection{Event 369} 
\begin{figure*}
    \caption{Chart of most interesting objects for Event 369 (equatorial coordinates, equinox 2000). The CVs, X-ray binaries and Z~And star as documented in our Section `Technical remarks'. The five enumerated objects are listed in Table~\ref{tab:allCVs}; the full list of objects is depicted in the charts in Fig.~\ref{fig:all369}. Highlighted is asterism `west wall of Zigong'.}
    \label{fig:369}
	\includegraphics[width=\textwidth]{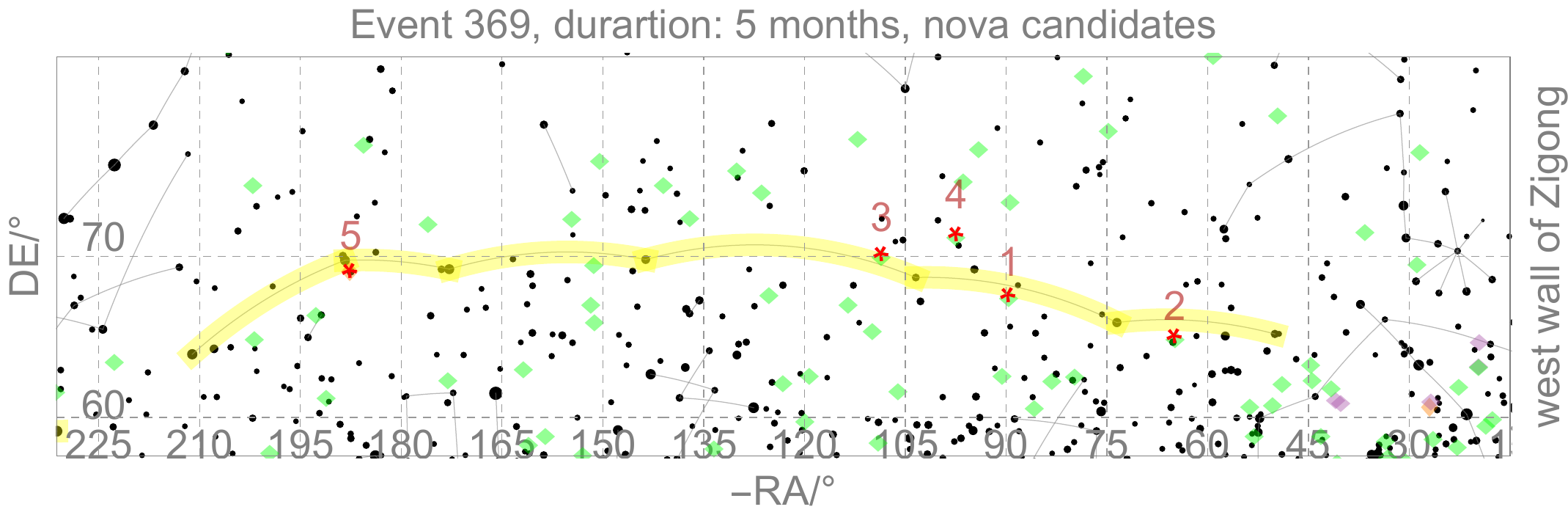} \\
\end{figure*}

 As the text describes the position of the object at the `west' wall of the Purple Palace, our star chart in Fig.~\ref{fig:369} displays only one line (yellow) representing this wall. A possible counterpart object could be almost anywhere in the chart but we consider it more likely being close to the asterism line (see record in Tab.~\ref{text:369}). As this area of the sky covers a large part of the circumpolar area far away from the Milky Way, the density of any type of object is low. 
 All SNRs and all PNe displayed in the map Fig.~\ref{fig:all369} are located in far galaxies and, therefore, not considered as candidates for the event and as well as the few PSRs. We should mention that the two almost circular PN A66~6 and PN HFG~1 are located at the end of the `west wall of Zigong'. The central star of PN HFG 1, V664 Cas, is a detached pre-cataclysmic binary \citep{shimanskii2004}, surrounded by a shell from the ejection of the common envelope, still visible as a planetary nebula. There is no astrophysical possibility of any recent eruption. Furthermore, an ancient Chinese astronomer would likely have reported an appearance at this position in the sky `at the northern end of the west wall' or `in space of the Guest Houses' or `guarding the Canopy of the Emperor' which would have been a more important astrological relevance. Thus, we consider these areas of the search field extremely unlikely related to the given guest star although they are correctly included in our map. Only the map of CVs (2nd in Fig.~\ref{fig:all369}) returns a dense coverage but, of course, only few of them are bright enough to flare up to naked eye magnitude and we note that only one of them, BZ~Cam, is next to a known PN candidate (see Fig.~\ref{fig:369}). 

 There is no hint in the text concerning the apparent magnitude of the appearance. As this is an area in the north, far away from any bright disturbances (such as Moon, planets, Milky Way), even a faint guest star (4~mag) could be recognized as `new' object. The record preserves that the object faded away after five months and, thus, it is a slow nova (type Nc) by definition implying a relatively small amplitude $A\leq10$~mag. Therefore, object~6 in our Tab.~\ref{tab:allCVs} (SDSS J100516.61+694136.5) is too faint at all for a naked eye nova candidate and not highlighted in Fig.~\ref{fig:369}. Only the brighter targets, i.\,e. three CVs and one symbiotic binary, deserve a closer look and are highlighted in Fig.~\ref{fig:369}. 
 
 Two candidates, LU~Cam and 1RXS J041924.8+653006 (object 1 and 2 in Fig.~\ref{fig:369}), need to brighten by 12 to 13~mag to reach 4~mag. These amplitudes would be possible for fast novae but exceed considerably those typical for slow novae, implying that these candidates are not very likely. Brighter are ASASJ071404+7004.3 and BZ~Cam, with average V magnitudes 12.2 and 13.3 mag, resp. For the first target (object 3 in Fig.~\ref{fig:369}), it is not yet certain that this is a CV: The ASAS-light curve of ASAS J071404+7004.3 does not show any outbursts but $\sim1.5$~mag eclipses and $\pm0.5$~mag flickering and its period of $\sim22$~d is neither typical for CVs nor for symbiotic stars. 

 \subsubsection{BZ Cam} 
 The last remaining CV candidate, BZ~Cam (object 4 in Fig.~\ref{fig:369}), is a nova-like and situated close to a nebula which is classified as PN candidate. BZ Cam shows a slow long-term variability during the past two decades, in the range V$=12.0-13.3$~mag, with two very brief excursions to V$\sim14.5$~mag, as typical for the VY~Scl sub-type of nova-like variables (see light curve in the LCG archive of AAVSO). Its orbital period of 3.68 hours places this star within the peak of the period distribution of classical novae between 3 and 4~hours, well known and recently confirmed by \citet[in prep.]{fuentesmorales2019}. If BZ~Cam erupted with a typical nova amplitude of 11 to 13~mag, it would have been really bright (0 to 2~mag) and a $t_3$ decline time of two or three months would be compatible with the long naked-eye visibility. 

\begin{figure*}
    \caption{Image of BZ Cam embedded in the red and partially bluish bow shock nebula EGB~4; the image is reproduced from \citet[p.\,3]{bond2018}. \textbf{Subfigures:} ($1$) The photograph (upper left) was obtained with the Kitt Peak National Observatory Mayall 4~m telescope on 1996 March 13. Red is H$\alpha$ and green is [O\,{\sc III}]. Height of frame is 6\farcm1; north is at the top and east on the left, and logarithmic stretches were used. It shows that the nebular structure is much bigger than seen on the images of surveys like DSS or PanSTARRS. The circles in subfigures~2, 3, and 4 (cyan, yellow, black-red) do not refer to circular shaped structures of filaments but they symbolise the curvature radius of these structures. Especially for the middle bow it is obvious that the bow does not perfectly fit a circle but a hook (subfigures 1 and 2) and the yellow circle (subfigure~3) is only an approximate shape: The structures in the nebula do not remain as exact circles while the binary moves like a bullet. ($2$) The innermost circle fits the greenish bow south of the star and covers the greenish region surrounding it. ($3$) The yellow circle (bottom left) is a fit to the innermost red bow (hook) northwest of the binary and ($4$) the black circle (bottom right) is a fit to an even bigger filament and the white dashed bows indicate further bows.}
    \label{fig:BZCam_draw}
    \includegraphics[width=.93\textwidth]{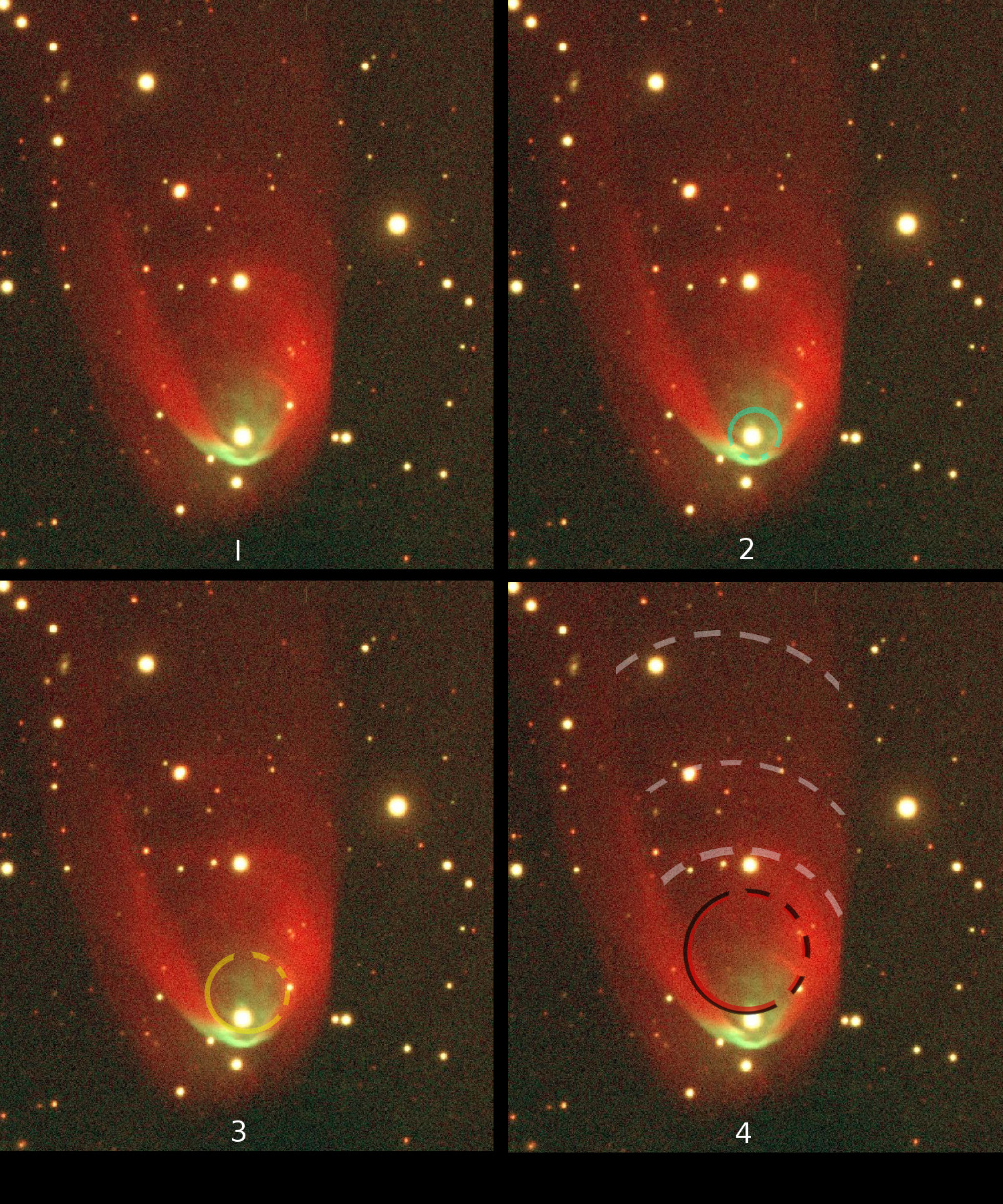} \\
\end{figure*}

 Of special interest is also the faint nebula EGB~4 next to BZ~Cam which seems to be structured (Fig.~\ref{fig:BZCam_draw}): There is a small $(16\pm3)\arcsec$ blue nebula exactly around the star (highlighted in subfigure~2).\footnote{The size estimate depends on the measurement: if we take the inner or outer rim and towards which direction.} Additionally, there is a tail-like red nebula that shows several shell-like structures with centres north of the star (subfigure~3, 4): In our Fig.~\ref{fig:BZCam_draw}, the bigger black circle (subfigure~4) has a curvature radius of $41\arcsec$ and is the lowest edge of a bow-shaped structure extending to $51\arcsec$ (white dashed). The smaller yellow circle (subfigure~3) which is attached to a bow-shaped filament within the red nebula has a curvature radius of $(27\pm2)\arcsec$. The nova-like star is situated at the edge of this red nebula (apparently caused by H$\alpha$ emission) and surrounded by the above mentioned small bluish [O\,{\sc iii}] nebula. This multiple shell structure is already described and discussed in \citet{griffith1995} and they even propose repeated eruptions as cause. The innermost parts of the nebula can already be seen on DSS and PanSTARRS survey but a deeper photography of the multi-shell structure was taken at the Kitt Peak 4~m telescope, is published in \citet{bond2018} and cited in our Fig.~\ref{fig:BZCam_draw}. This figure strengthens the suggestion to interpret the many bow-shapes (dashed white) as remnants of recurrent novae. Additionally, the image shows that the circles we applied in Fig.\,\ref{fig:BZCam_draw} to estimate the age of those eruptions, are very rough fits to the bow-shaped filaments within the elongated tail-like structure formed by the movement of the binary.
%

 With a Gaia parallax of $2.6882\pm0.0374$~mas \citep{brown} the radius values of these three nebulae amount to 1.5, 2.7, and 4.2~pc respectively. Applying typical nova shell expansion rates \citep{valle1997} between 500 and 1200~km/s, this leads to kinematic ages of the order of 2, 4, and 6 millennia respectively for the inner three nebulae. With the outmost of them, we leave historical timescales, i.\,e. observations of the eruption that caused the third, forth, and fifth bow cannot be preserved in written human cultures. 
 
 According to Gaia proper motion of BZ Cam (RA,DE)$=(-2.233,-28.677)$~mas/yr, the star's displacement is $\sim29\arcsec$ per millennium towards south, compatible with a possible nova ejection of all three nebulae by BZ Cam, about 2, 5 and 8 millennia ago. With regard to the sometimes big error bars of Gaia parallaxes for binary stars and including a possible proper motion of the shell(s), these two age estimates are compatible. 
 
  This suggests BZ~Cam to be the first known candidate for a Nr-type nova with a recurrence cycle of $\sim2000$ years. This suggestion is supported by the large tail shown in \citet{bond2018} (reproduced in our Fig.~\ref{fig:BZCam_draw}): The bow filaments within it have curvature radii of 16, 27, 41--51, 61, 90.5 seconds of arc, all $\pm3$\arcsec. This series seems to fit a pattern of multiples of 15 with one irregularity (41--51 instead of 45 at a relatively broad structure which could equally likely be remnant of one or a few eruptions) implying periodic eruptions. If the last eruption could have been observed by ancient Chinese astronomers 1650 years ago and if the period of eruptions is something like 2 millennia, we expect a new eruption in $\sim500$~years. However, with regard to the unknown expansion rates and the uncertainty of the proper motion, new measurements are required for a proper prediction. The structure of the nebula only suggests a recurrent nova on millennium timescale. 

\subsubsection{CQ Dra $=$ 4 Dra} 
 In our search field, there is another interesting object: The symbiotic star CQ~Dra (object 5 in Fig.~\ref{fig:369}) with a usual $V$ range of $4.90-5.12$~mag already has naked eye visibility but the 5~mag star is $0\fdg5$ away from the 4~mag star $\kappa$~Dra. If CQ~Dra erupts as nova, it would certainly be visible for several months by ancient observers and it would be as bright as Sirius or even as Venus. In an area without such bright stars this would be very eye-catching, so this appears an exceptionally good candidate to explain the sighting. 

 \citet{reimers1988} presented CQ Dra as a possible triple star consisting of a cataclysmic binary as primary component and a red giant star as secondary in a wide eccentric orbit ($e = 0.30, P_\textrm{orb} = 1703$~d). Based on IUE observations, these authors suggest for this hypothetic CV an orbital period of 0.1656~d, again near the peak of the overall nova period distribution. However, \citet{wheatley2003} exclude the presence of a magnetic CV within the triple system due to the lack of a strong X-ray flux but accept either a non-magnetic CV or a single white dwarf as primary component in this system. This object will need further investigation, resolving the uncertainty whether there is a CV within the CQ~Dra system or not. In either case, it could be an interesting alternative counterpart for the event~369.   
 
 If the system that already has $V\sim5$~mag flares up by $A\approx10$~mag as a slow nova and reaches $-5$~mag it would be a rather spectacular appearance which could be visible even in daylight. In this case, we would expect more attention to the sighting than only one brief note and it should have been seen by all astronomers of the northern hemisphere and not only by those at the Chinese court. However, from this epoch (Late Antiquity) records of astronomical observation are rare in all cultures and even the medieval supernovae 1006 and 1054 are hardly recorded in Europe \citep[chap.~9 and 8]{stephGreen}. Additionally, it is not at all certain that a nova from CQ~Dra really reached daylight visibility: An amplitude of 10~mag or even 11~mag (V445 Pup$=$Nova Pup\,2000, cf. \citet{goranskij2010}) is possible but not the typical case for slow novae. More likely are amplitudes of $\sim7$~mag \citep{hoffmannVogtLux} which would make this star only a bright nighttime appearance.
 
 Thus, we consider an eruption of CQ~Dra as likely as an outburst of BZ~Cam. 

 \subsection{Event 386} 
\begin{figure*}
    \caption{Chart of most interesting objects for Event 386 (equatorial coordinates, equinox 2000). The suggested SNRs, G011.2-01.1 \citep[p.\,182]{stephGreen}, G011.2-00.3, and G007.7-03.7 \citep{zhou2018}, are marked with red six-pointed stars, and CVs, X-ray binaries and Z~And star as documented in our Section `Technical remarks'. Highlighted is asterism Nandou.}
    \label{fig:386}
    \includegraphics[width=.98\columnwidth]{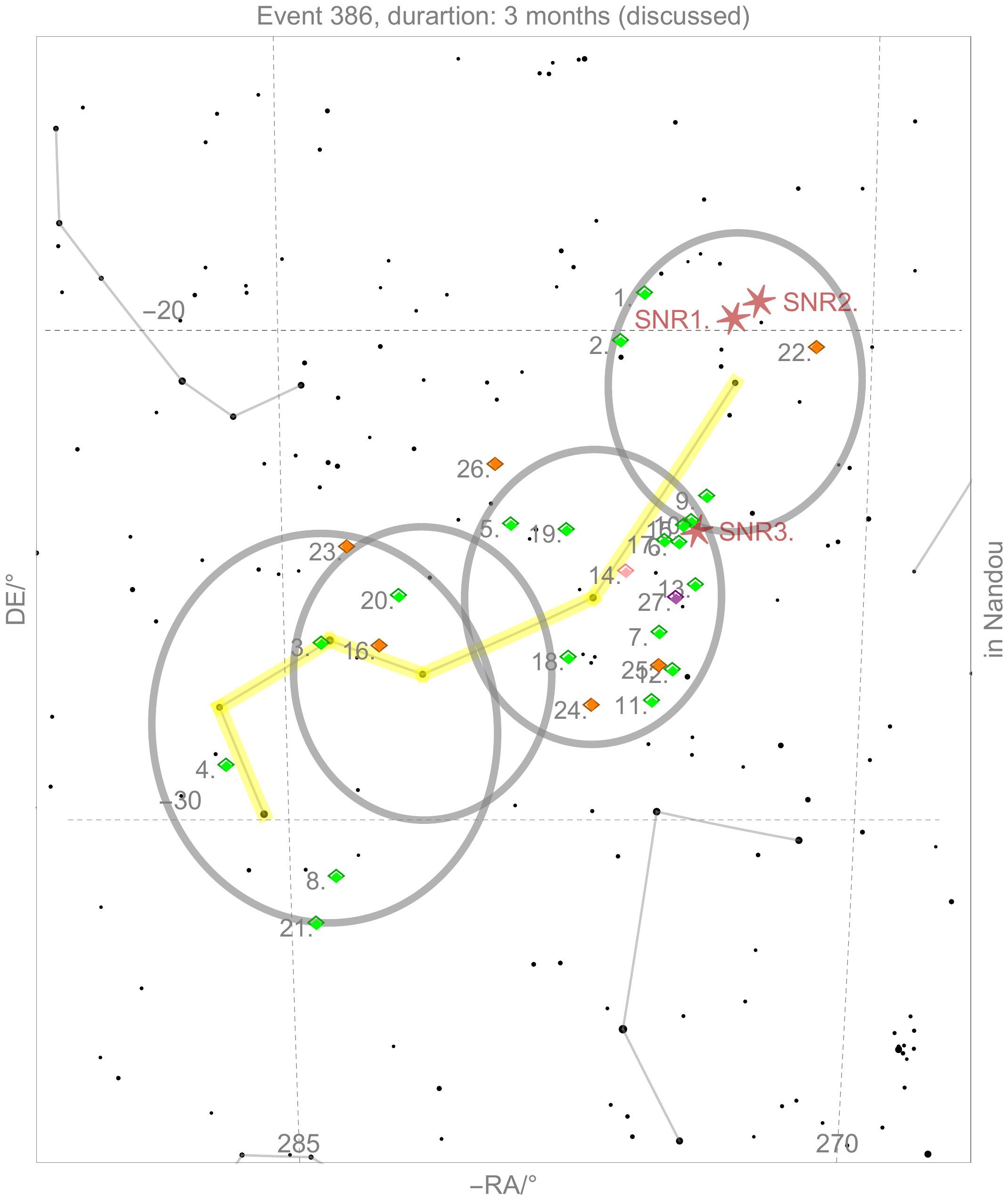} \\
\end{figure*}

 In contrast to the first sighting, this appearance is reported in an area close to the Galactic plane. The asterism Nandou, the Southern Dipper, is part of the `teapot' seen in the brightest stars of the IAU-constellation Sgr; its tip touches the area of high object density. The historical record does not give the position more precise than within the asterism. Thus, we need to apply our method to cover the constellation area with circles and search for objects in them. Fig.~\ref{fig:386} displays the supernova remnants, PN candidates, and CVs in the field which passed our criteria to potentially have flared up in historical time. The full set of objects of each type is displayed in the same map in the Online-Only Appendix and for a first glance in Fig.~\ref{fig:all386}. The CV candidates are listed in Tab.~\ref{tab:allCVs}.

 \subsubsection{Supernova suggestions} 
 In this case, \citet{stephGreen} consider their suggestions of SNRs as somehow likely but not certain. It seems to roughly fit the age but the historical description is so brief that certainty cannot be achieved. However, the SNRs in the middle of our search field G008.7-05.0 and G007.7-03.7 have small radio sizes of $26\arcmin$ and $22\arcmin$, respectively. G008.7-05.0 has not yet been suggested for a historical supernova but its position matches the description and the object has been observed several times with Fermi Large Area Telescope (LAT) since 2010, and is included in the LAT Supernova Catalog \citep[and references therein]{acero2016}. While \citet{stephGreen} suggest SNR G011.2-00.3 (`SNR1' in our map) at the upper end of the handle of the Dipper, \citet{zhou2018} suggest SNR G007.7-03.7 (`SNR3' in our map) as also possible after X-ray observations. This 22\arcmin-shell fits better the description because the text in Tab.~\ref{text:386} says `in the constellation of the Southern Dipper' and not its edge or north of it. Additionally, this shell is close enough that the supernova could achieve naked eye visibility and they estimate the age of the SNR being 500 to 2200~years. On the other hand, they also doubt on their suggestion with regard to the brief duration of the event which is reported for only three or four months. Thus, they conclude in case of a match of this SNR and the reported guest star that we deal with a low luminosity supernova.  

 \subsubsection{Discussion of alternative candidates} Cross-checking this area for CVs and symbiotic stars, we obtain the list of bright CVs with a total of 28 entries in Tab.~\ref{tab:allCVs} which are (theoretically) bright enough to become visible for the naked eye. With regard to the position of the search field, close to the horizon and next to the bright clouds of the Milky Way, we expect any naked eye discovery to be rather bright: Although a 5~mag object could be visible, a transient of this peak brightness would not be recognizable as `new star' in this field and a 4~mag transient would also be unlikely discovered, cf. \citet[Fig.\,3]{hoffmannVogtProtte}. We assume this guest star to have had a peak brightness of at least 2 or 3~mag. Fading to invisibility ($\sim6$~mag) within 3~months implies a slow nova but if the object was really bright (Venus brightness) it could also have been a moderately fast or fast classical nova with $t_2=26-80$ or $11-15$~days by definition. 
 
 Our search for X-ray binaries returned for instance the object V4641 Sgr which is enrolled in the VSX with a maximum brightness of 9.1~mag in $V$ and, thus, listed in our Tab.~\ref{tab:allCVs}. As this is a HMXB it cannot show nova behaviour and needs a special treatment: Apparently, this is an accreting black hole which attracted attention by an outburst -- but with regard to the AAVSO light curve the maximum was only 12.8~mag. Either way, as this is not a periodical behaviour but unpredictable, and it would likely not have a duration of several months. Thus, we neglect it as counterpart of the historical guest star sighting. 
 
 The symbiotic star V5569 Sgr is classified EA+BE+ZAND, its primary is probably a Be star, not a white dwarf, excluding any nova behaviour. 

 IGR J18245-2452 is placed near the core of the globular cluster M28 and refers to an X-ray transient and type-I X-ray burster, whose X-ray outburst of April 2013 was also observed by the Hubble Space Telescope as brightening in the visual band, reaching about 12~mag, however, varying in quiescence between 20.6 and 23.2 mag \citep{2013ATel.5003....1P}. This object can definitively be disregarded.

 \subsubsection{Nova and CV-related suggestions} There are four relatively bright Z~And stars: V5759 Sgr, AS 327 and V1988 Sgr (all with mean V$\sim13.0$~mag) as well as V2601 Sgr (V$\sim14.6$~mag), each of them could be considered as a possible candidate for a nova. 

 Among the CV-related stars in this field there is an outstanding candidate: the recurrent nova V3890 Sgr with three recorded nova eruptions in 1962, 1990 and 2019 \citep{darnley2019, munari2019} and had been observed in all bands from radio (e.\,g. ATel $\sharp$13047, ATel $\sharp$13050, ATel $\sharp$13089) to X-ray \citep[and ATel $\sharp$13124]{orio2020} and gamma ray (ATel $\sharp$13114). During the last eruption the star was observed at $V=6.7$~mag near eruption maximum (Pereira, 2019: vsnet-alert 23505) but the rise phase was not covered, so it is well possible that its true maximum brightness was near $V \sim6.0$~mag, at the limit of naked eye visibility. Its orbital period of 519.7~d places this star into the RS~Oph group of recurrent novae, characterized by a late-type giant star as secondary component \citep{darnley2019}. Apparently, V3890~Sgr is a classical recurrent nova, like the Nr candidate KT~Eri, suggested as counterpart for the event 1431 by us \citep{hovoMNRAS2020}. 

 Among the remaining targets we would like to mention some additional relatively bright candidates. V1223 Sgr, a DQ~Her type cataclysmic variable with an orbital period of 3.37 hours (again near the peak of the nova period distribution), was observed between 12.5 and 13.8~mag during last decades (see LCG archive of AAVSO) showing slow irregular variability.

 Overall, the identification of the progenitor of this guest star remains uncertain. 

 \subsection{Event 393} 
\begin{figure*}
    \caption{Chart of most interesting objects for Event 393 (equatorial coordinates, equinox 2000). The suggested SNR G347.3-00.5 \citep{wang1997} is marked with red six-pointed stars, and CVs, X-ray binaries and Z~And star as documented in our Section `Technical remarks'. Highlighted is asterism Wei [LM6].}
    \label{fig:393}
    \includegraphics[width=.95\columnwidth]{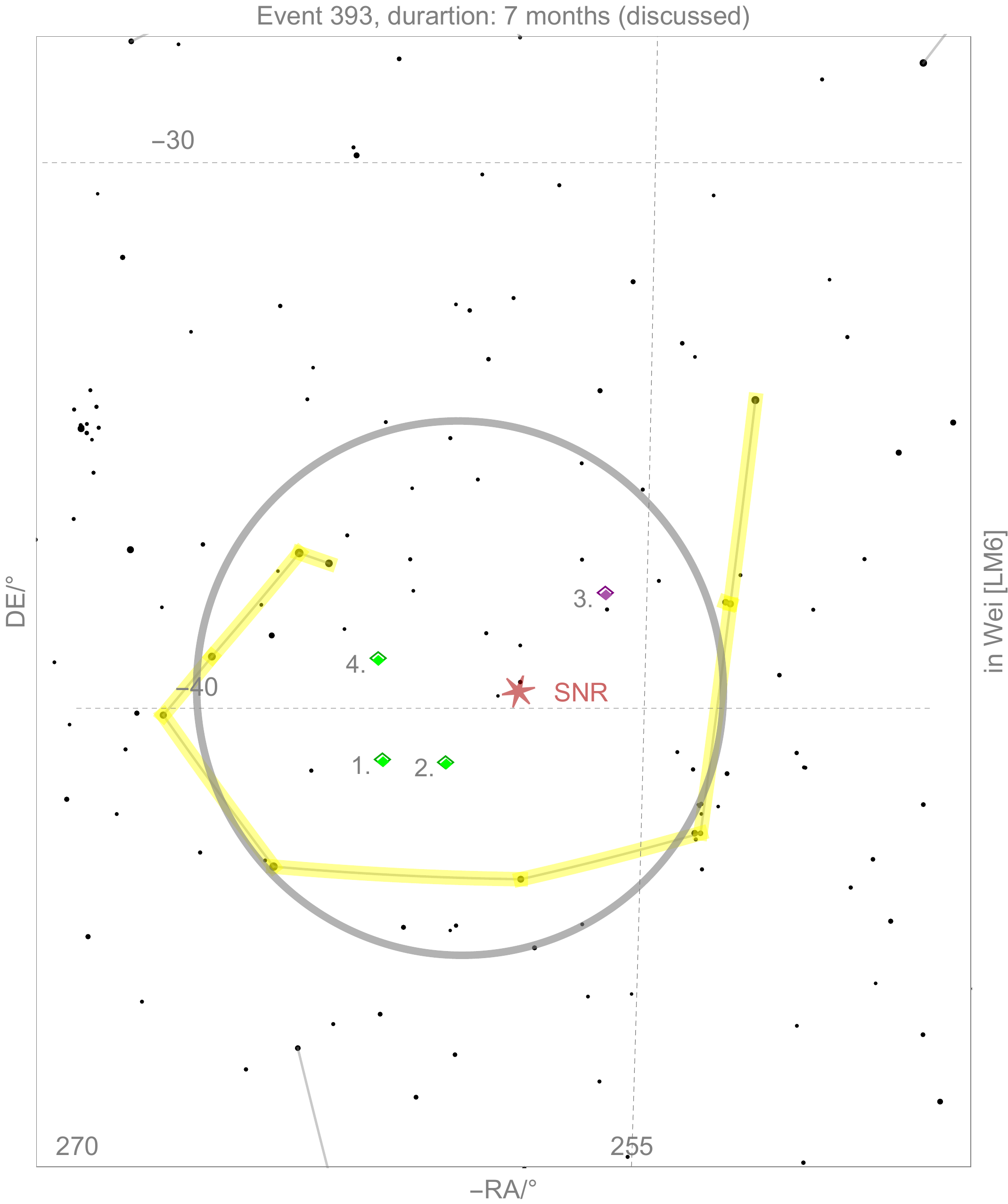} \\
\end{figure*}

 In 393 a guest star was observed in the asterism of Wei in Sco. It is reported `in the middle of Wei' and as this `Wei' is shaped like a container we expect the object having appeared really close to the centre of a circle which could be inscribed into this container shape. Fig.~\ref{fig:393} displays the three maps of our candidate search, and Tab.~\ref{tab:allCVs} lists the CV candidates. 
 
 \subsubsection{Supernova suggestion:} In this case, \citet{stephGreen} have doubts on the suggested identification of the SNR G347.3-00.5 \citep{wang1997}; they cite and comment on it but resume that there are always alternatives. Due to the many SNRs in this field (they count 14, our more recent map shows already c.\,20) it might be always uncertain. However, most of the SNRs are too old which, on the other hand, makes the identification somehow likely depending on the determination of the distance and, thus, the kinematic age. The suggested SNR is really `in the middle of Wei' and, therefore, matches the given position perfectly. However, considering the event~393 as (slow) nova is a valid option, the result of our search for nova candidates is given in Tab.~\ref{tab:allCVs}. 
 
 \subsubsection{Discussion of alternative candidates} 
  The search for X-ray binaries returned V884~Sco, a bright (V$\sim6.5$ to 6.6~mag) high mass X-ray binary (HMXB) and, therefore, not an accreting compact object but only ellipsoidal giants. It consists of two O-type stars with a variability feature which is caused by the deformation of the components although their Roche lobe is not filled and, thus, there is no possibility for nova behaviour. The GCVS Team mentions a surrounding ring nebula, Sh~2, but this probably refers to a huge hydrogen cloud. Thus, we neglect this object in our consideration of potential candidates. 
\begin{figure}
    \caption{IGR J17195-4100 is a polar (DQ Her-type) in Sco which is 8\farcm76 separated from a nebula detected in the infrared. The images made with CDS Aladin show the comparison of the IR images of 2MASS and Vista (JYZ) Survey (left) and the comparison of the Chandra X-ray and the Vista-IR image (right). The target is in the very center where the X-ray emission is registered.}
    \label{fig:IGR}
	\includegraphics[width=.48\columnwidth]{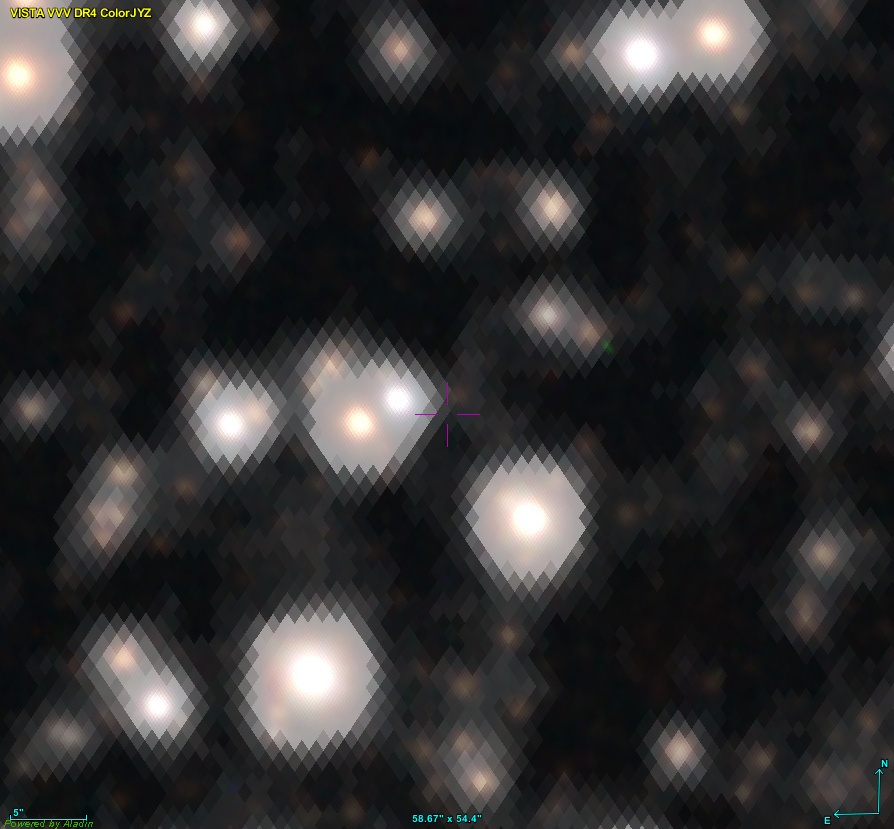} \hfill
	\includegraphics[width=.48\columnwidth]{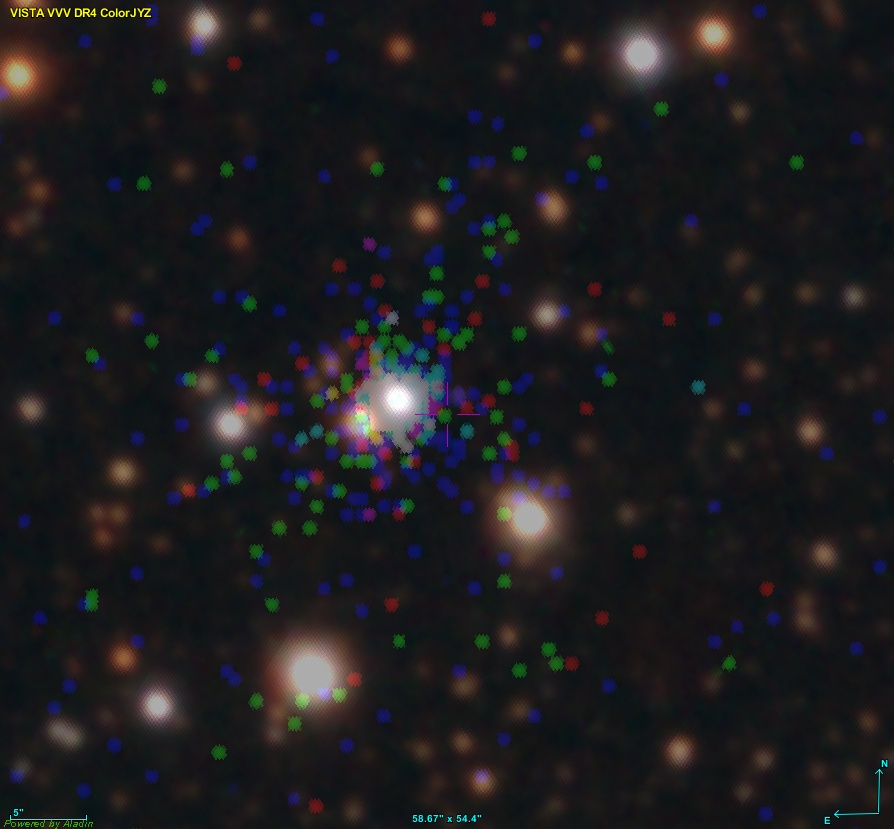}
\end{figure}
 
 \subsubsection{Nova and CV-related suggestions} The first target is V902 Sco (Nova Scorpii 1949). \citet{duerbeck1987} reports an uncertain identification within a `group of stars in an obscured region of the Galaxy' with a possible minimum brightness of $\sim20$~mag and $t_3 = 200$~d. Since V902 Sco was $\sim11$~mag at maximum and apparently a slow nova, it is rather unlikely that it ever had been reached naked eye visibility. 
 
 There are two cataclysmic variables almost in the middle of the Queue (Wei). The first is IGR J17195-4100, an intermediate polar (DQ Her-type). It is especially interesting because it has a small angular separation to the nebula [MHL2007] G347.061902.0167 1 which is classified as possible PN \citep{steene1993} but could also be a young stellar object, detected in mid-inferred \citep{mottram2007}. We checked all available IR, X-ray, and optical images with CDS Aladin in this crowded field and found only a very strong radio source and a faint IR-bridge between the stars in this group (Fig.~\ref{fig:IGR}). Yet, because of the high object density no relation is obvious at first glance.  

 The second CV in the middle of the Queue (Wei) is V643 Sco, about $\sim2$~mag brighter than IGR J17195-4100. It is a dwarf nova of Z~Cam sub-type varying between 13.0 and 14.2~mag. With regard to the long visibility of the guest star (7 months) and its proximity to the horizon and the Milky Way, this event should have been a rather bright appearance. Once discovered, a bright object could be traced for a long time but with a bright background and a vicinity of many bright stars a faint `new star' would unlikely be to recognize, cf. \citet[Fig.\,3]{hoffmannVogtProtte}. A classical nova with amplitudes of 11 to 13 mag could brighten the star reaching naked eye visibility. Even a slow nova cannot be excluded as possible counterpart as there are 13 Nc-type (very slowly declining) novae listed in the VSX with reliable maximum and minimum magnitudes reveal an average amplitude of 7.5~mag (extrema: 4.3~mag for V1825 Aql and 12.5~mag for V445 Pup). 

 V643~Sco and IGR J17195-4100 are both possible candidates and of same likelihood. 

\section{Conclusion} 
  An overview mentioning the most likely identifications is given in Tab.~\ref{tab:ergebn}. The first of the events (369) is likely a nova. For the other two events (386 and 393) it is currently undecidable what caused them.  
 \begin{table} 
  \caption{The most important candidates for modern counterparts of the historical guest stars. Our new suggestions are highlighted in bold face. In the case of nova candidates, we mention here only the names of the most probable identifications, which seems to be viable alternatives to SNRs, hitherto mainly suggested by previous authors.} 
  \label{tab:ergebn} 
  \begin{tabular} {c|p{.42\columnwidth}p{.42\columnwidth}} 
  year & SNRs & Nova candidates \\ 
  \hline 
  369 & no & \textbf{BZ Cam}, a nova-like CV as possible recurrent nova  \\ 
		&& \textbf{CQ Dra} as nova, a symbiotic binary or a triple containing a CV\\
		&& and 2 other CVs\\
		\hline 
  386 & G011.2-01.1 \citep[p.\,182]{stephGreen} & \textbf{V1223 Sgr}, intermediate polar \\ 
	  & G011.2-00.3 & \textbf{V3890 Sgr}, a known recurrent nova,\\
	  & G007.7-03.7 \citep{zhou2018} & 4 further symbiotic binaries\\
	  & \textbf{G008.7-05.0 (new)} &   \\	  
  \hline 
  393 & G347.3-00.5 \citep{wang1997} & \textbf{V643 Sco}, a Z~Cam-type dwarf nova;\\
		&& \textbf{IGR J17195-4100}, intermediate polar \\
  \end{tabular}
 \end{table}

 \textbf{Event 369.} Among the plenty of CVs in the field of event 369, four fit perfectly the description in the ancient text and are bright enough to reach naked eye visibility in case of a classical nova eruption. LU~Cam, for instance, fits the given position but with a G mag of 16.4 is a not very likely candidate for a slow nova. In contrast, BZ~Cam is a very interesting candidate for a slow nova, especially because its shell's structure could witness recurrent explosions $\sim2$ to $\sim6$~kyr ago. 

 CQ~Dra is the brightest target in our search field, a multiple system of 4~mag in $G$ (naked eye visibility). It could be a symbiotic binary or a triple system with a cataclysmic binary component. Either way, a nova in this naked eye system will definitely have naked eye visibility. If this system produced a nova with an amplitude of 5 to 7~mag (which would be typical) it would be a rather good candidate to explain the sighting. However, there are also novae known with amplitudes up to 10 or 11~mag (or even more) and in this case, the system could have reached daylight visibility. In case of such a spectacular event, one could expect more attention to have paid to it in records of this and other astronomical cultures. 
 
 Apparently, the sighting of 369 could only refer to a classical or recurrent nova and not to a supernova. 

 \textbf{Event 386.} For this event, there had been three suggestions of SNRs among which the most recent one of the radio and X-ray source G007.7-03.7 is considered most likely \citep{zhou2018}, and we added the suggestion of SNR G008.7-05.0. However, there are also some 15 old pulsars in this area, one symbiotic and one X-ray binary, as well as $\geq20$ bright CVs. With the given conditions (bright peak and $t_3\simeq2$ or 3~months), we found only one valid CV candidate (V1223~Sgr) and five symbiotic binaries for classical novae: V1988 Sgr, V5759 Sgr, AS 327, V3929 Sgr, and our favourit V3890 Sgr, a recurrent nova with three recorded modern nova eruptions, reaching nearly naked eye brightness during the last one in 2019. At the last outburst, the brightest record reported 6.7 mag but possibly missed the real peak and the peak brightnesses varies from outburst to outburst by $1-2$~mag.
  
Further investigation and follow-up observations are required to ultimately decide what caused this ancient `guest star' but with regard to its 3~months duration a classical nova appears more likely than a supernova. 
 
 For \textbf{event 393} one SNR has been suggested: G347.3-00.5 \citep{wang1997} and considered `probable' by \citet{stephGreen} and the U Manitoba catalogue \citep{manitoba}. Alternative identifications are the dwarf nova V643 Sco and the DQ Her-type magnetic cataclysmic system IGRJ17195-4100. All three mentioned targets have similar likelihood for this event.

 \textbf{Summarising the whole study}, we conclude that there is no conclusive evidence that one of the three \textit{longue-durée} guest stars of the 4th century have been supernovae. Neither there is evidence that they have been novae. With regard to the ambitions of some scholars to quickly interpret a certain historical guest star either as a nova or as a supernova, this study is unveiling the indefiniteness of possibilities offered by nature to interpret these appearances. On the one hand, this might teach caution but on the other hand, ancient guest stars are the only possible way to augment timescales of variable star research significantly beyond our current knowledge because telescope observations are limited to $\sim10^2$~years back. Therefore, we believe that it is worthwhile the effort.  
 
 \section*{Technical remarks} 
\begin{figure}
	\includegraphics[width=\columnwidth]{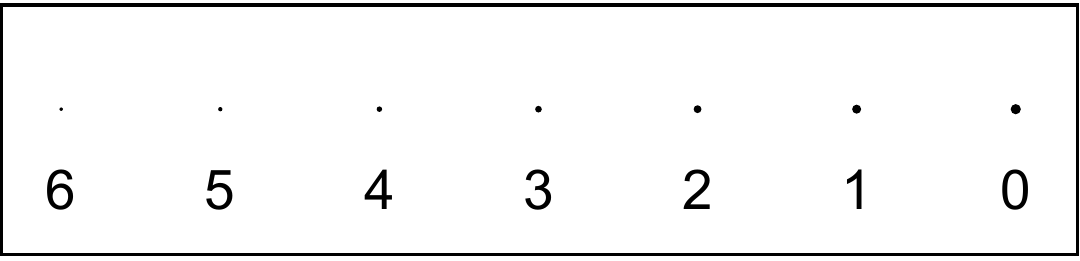} 
    \caption{Scale of star sizes for the maps in the appendix.}
    \label{fig:magScale}
 \end{figure}

  All maps use the same symbolism: They display the stars of the Yale Bright Star Catalog (HR), equatorial coordinates, equinox 2000, scaled according to their magnitude (Fig.~\ref{fig:magScale}). 
 
 The stick figure lines of the Chinese asterims are taken from Stellarium 0.19.2 (version by contributed by Karrie Berglund, Digitalis Education Solutions, Inc. based on Hong Kong Space Museum star maps), slightly changed in cases where the identified star was not in the Yale Bright Star Catalog (HR). This version contains a few more asterisms than the Suzhou map which is why our search circles are developed by always cross checking with the original historical map (as described in our earlier publications).

 The planetary nebulae (PN) are indicated with a ring with central point $\odot$, PN candidates with $\otimes$, and all types of binary stars with diamond $\bDiamond$ but the bright ones are highlighted by a filled diamond $\bLozenge$. The SNR are symbolized with a red `O', the PSR with blue $\ast$. The catalogue of PN and PN candidates originates from Simbad, downloaded in April 2020, while the catalogues of CV, X-ray binaries, and Symbiotic stars are downloaded from the VSX catalogue of the AAVSO (April 2020). In these maps, the nebulae are mapped in blue while there nearby binaries are mapped in orange (symbiotic stars), purple (X-ray binaries) or pink (LMXB), and green (CVs).

\section*{Acknowledgements}
 This research has made use of `Aladin sky atlas' developed at CDS, Strasbourg Observatory, France \citep{aladin2000,aladin2014}. Thankfully we made use of the VSX variable star catalogue of the American Association of Variable star Observers (AAVSO) and of the SIMBAD data base (CDS Strassbourg) \citep{wenger2000} and the ATNF Pulsar Catalogue \citep{atnf} and of Stellarium \url{http://stellarium.org}. S.H. thanks the Free State of Thuringia for financing the project at the Friedrich Schiller University of Jena, Germany. N.V. acknowledges financial support from FONDECYT regularNo. 1170566 and from Centro de Astrofísica, Universidad de Valparaíso, Chile. We thank the anonymous referee for his benevolent review and many constructive advise to improve the readability of the paper! Ralph Neuhäuser (AIU, Friedrich-Schiller-Universität Jena) had the initiative and idea to reconsider historical nova identifications including new nova candidates in a transdisciplinary project. We thank him to have brought us together.
 


\bibliographystyle{mnras}
\renewcommand{\refname}{\bf PAPERS of our SERIES}

\renewcommand{\refname}{\bf REFERENCES}
\bibliography{altNovaeA} 



\appendix
\definecolor{darkred}{rgb}{.7,0,0} 
\definecolor{pink}{rgb}{.9,.7,.7} 
\section{The maps which we used to generate the list of potential candidates.}
In addition to these maps of objects in our search fields, the according interactive CDF files are published in the Online-Only part of the paper. 

 \begin{figure*}
    \caption{Charts for Event 369 (equatorial coordinates, equinox 2000). The map supernova remnants (SNRs: `$O$') and pulsars (PSRs: $\ast$) (top) only shows some extragalactic SNRs, in the galaxies M82, IC\,342, and NGC\,1569. The map of planetary nebulae (PNs: $\odot,\otimes$, 3rd map) also shows plenty of PNe in M82 and IC\,342; the most interesting objects are the two `PN candidates' ($\otimes$) in the middle of the asterism. The CVs $\diamondsuit$ are displayed in the 2nd map, those of them brighter than 18~mag and the apparent close pairs of nebulae and CVs are displayed in the fourth map. The fifth map has the most interesting objects highlighted: see discussion.}
    \label{fig:all369}
	\includegraphics[width=1.6\columnwidth]{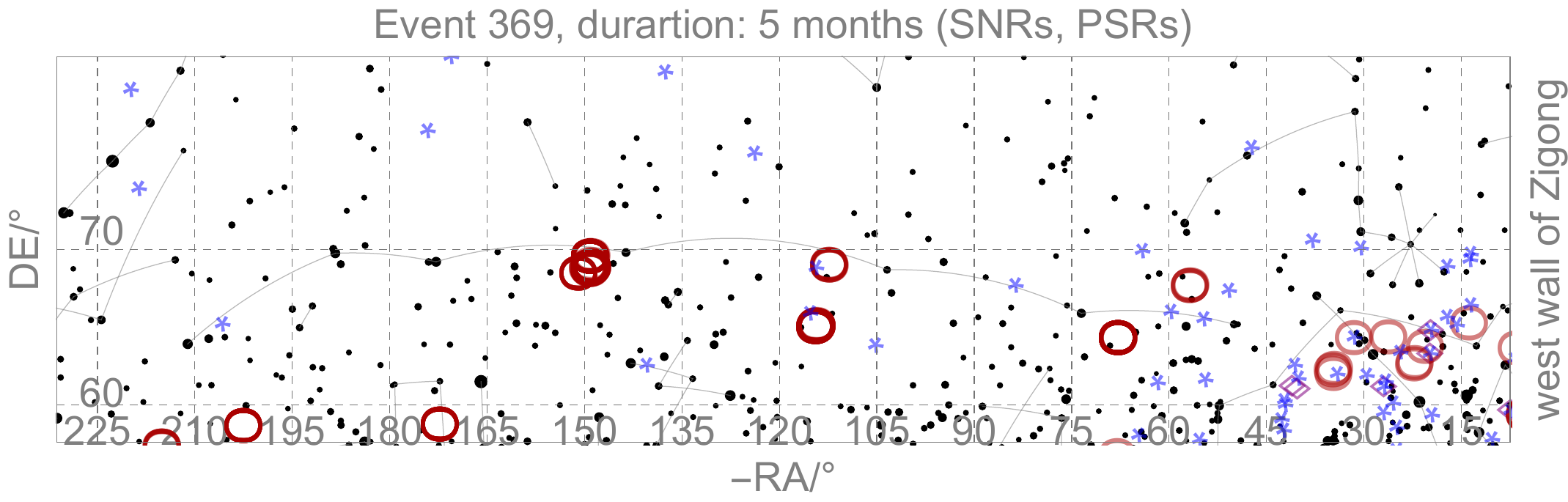} \\
	\includegraphics[width=1.6\columnwidth]{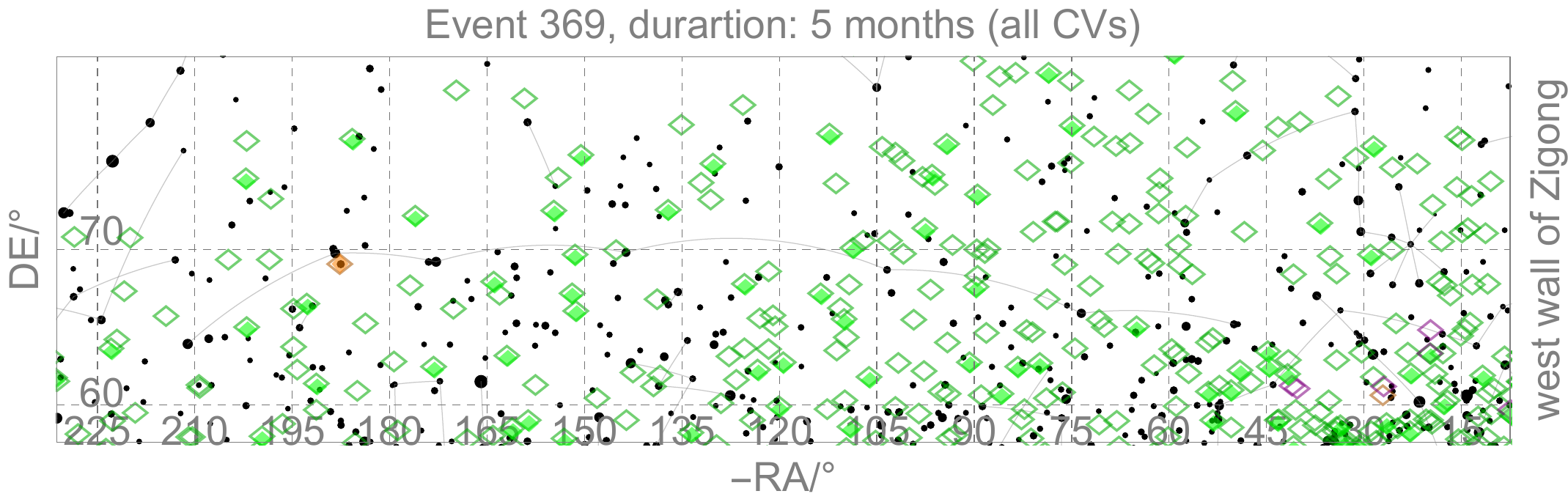} \\
	\includegraphics[width=1.6\columnwidth]{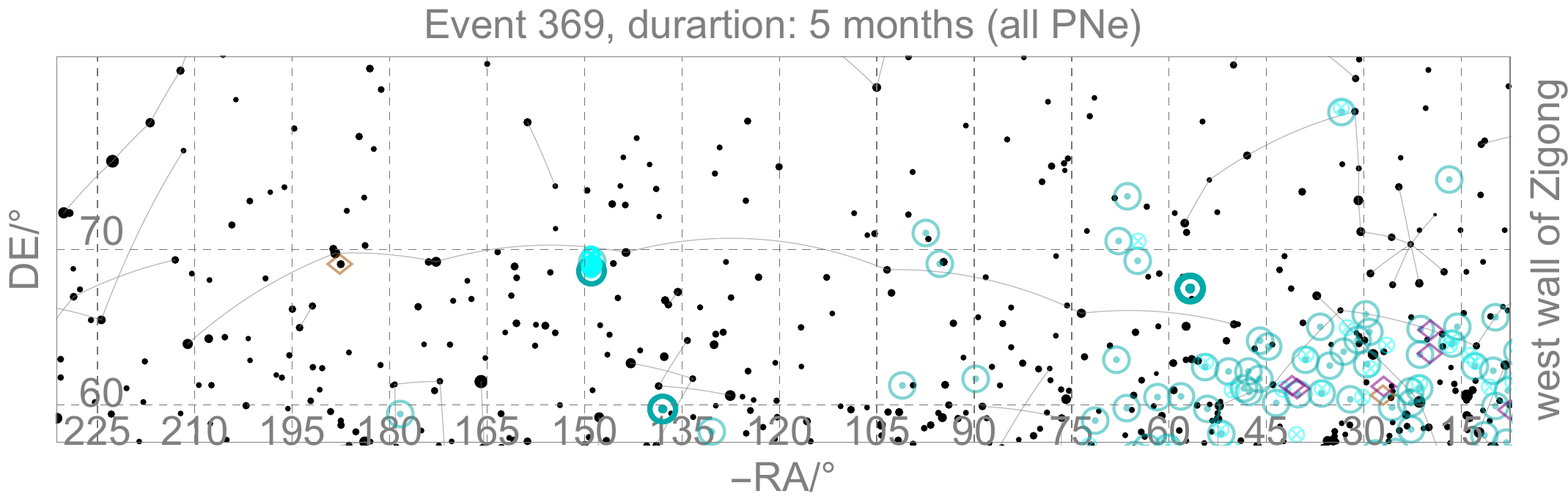} \\
	\includegraphics[width=1.6\columnwidth]{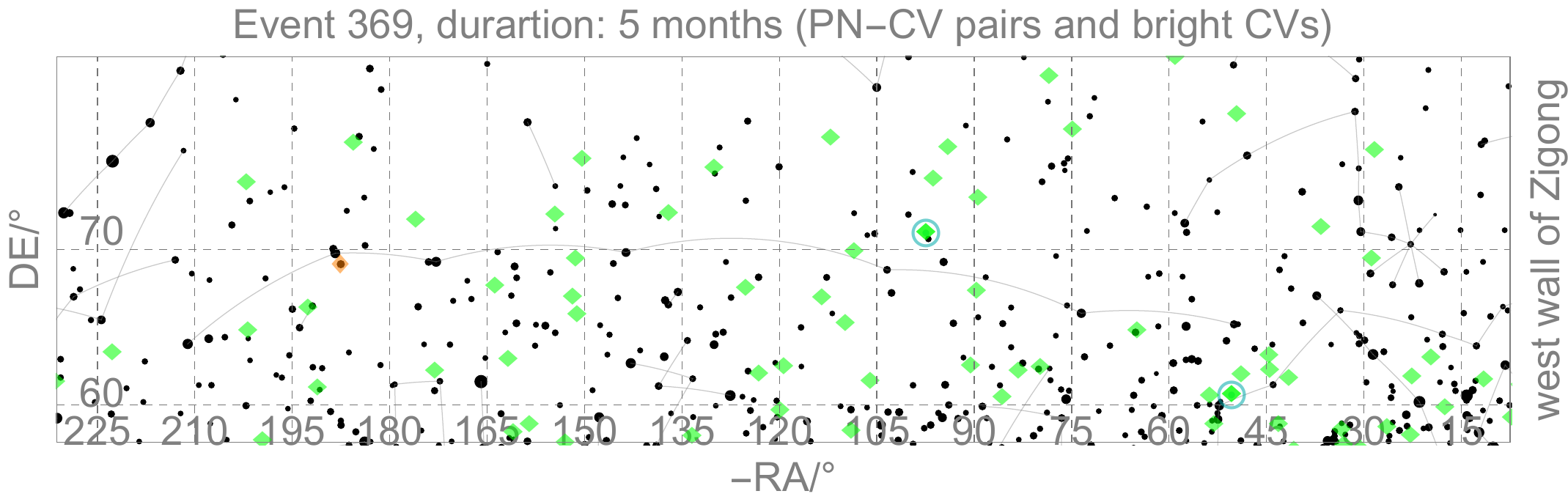} \\ 
	\includegraphics[width=1.6\columnwidth]{cvMap369sel.pdf} \\
\end{figure*}

 \begin{figure*}
    \caption{Charts for Event 386 (equatorial coordinates, equinox 2000). The four maps display the SNRs `$O$' and PSRs $\ast$ (upper left), the PNe $\odot$ and PN candidates $\otimes$ (upper right), the CVs $\diamondsuit$ and the brightest of them highlighted by filling of the shape (lower left), and the planetary nebulae close to binaries (lower right). The suggested SNRs, G011.2-01.1 \citep[p.\,182]{stephGreen}, G011.2-00.3, and G007.7-03.7 \citep{zhou2018}, are marked with red six-pointed stars  (top left of the asterism). The grey circles are covering the search field in which we suggest to look for a possible counterpart. Their coordinates are displayed in Tab.~\ref{tab:fields}.}
    \label{fig:all386}
    \includegraphics[width=.98\columnwidth]{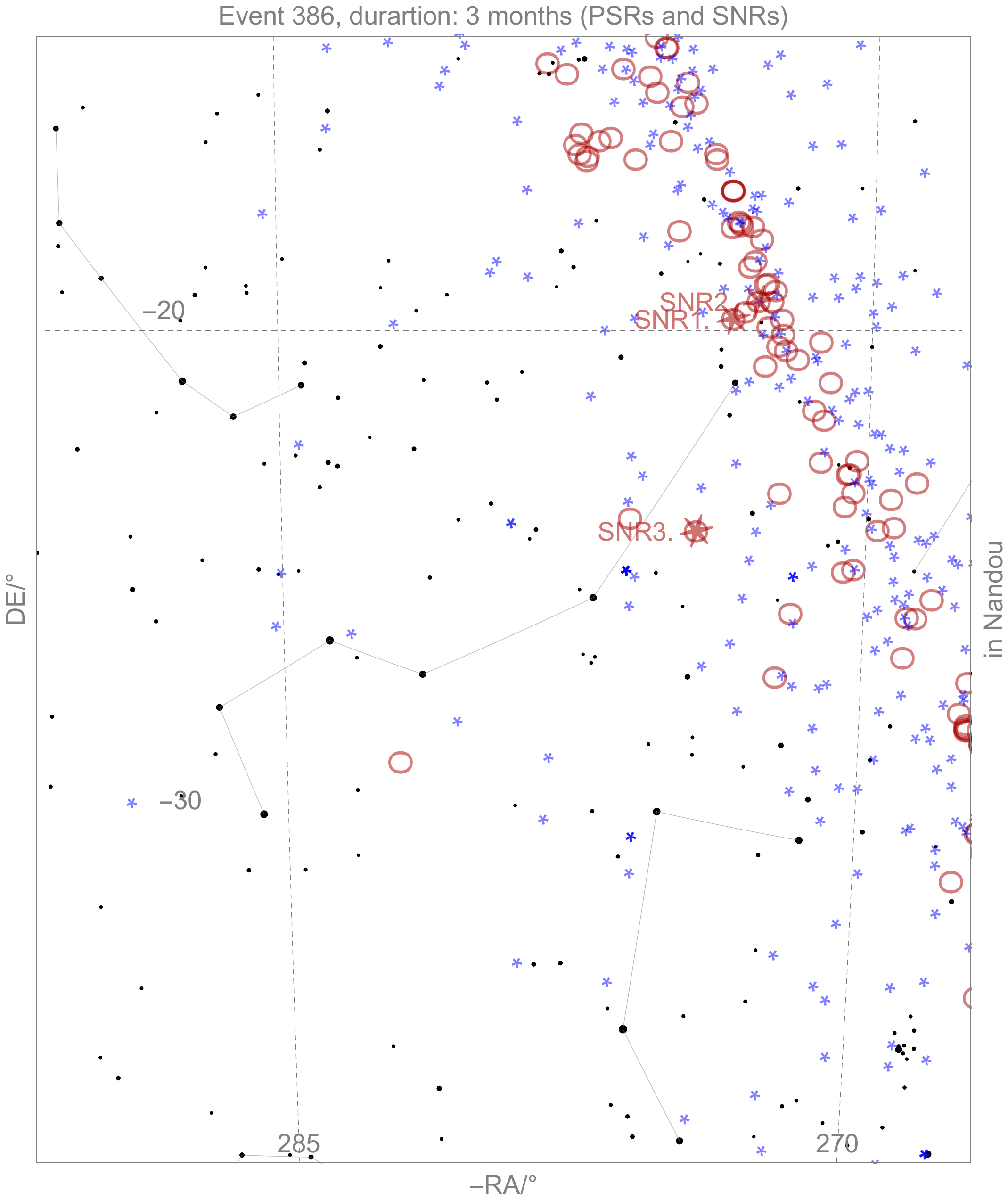} \hfill
	\includegraphics[width=.98\columnwidth]{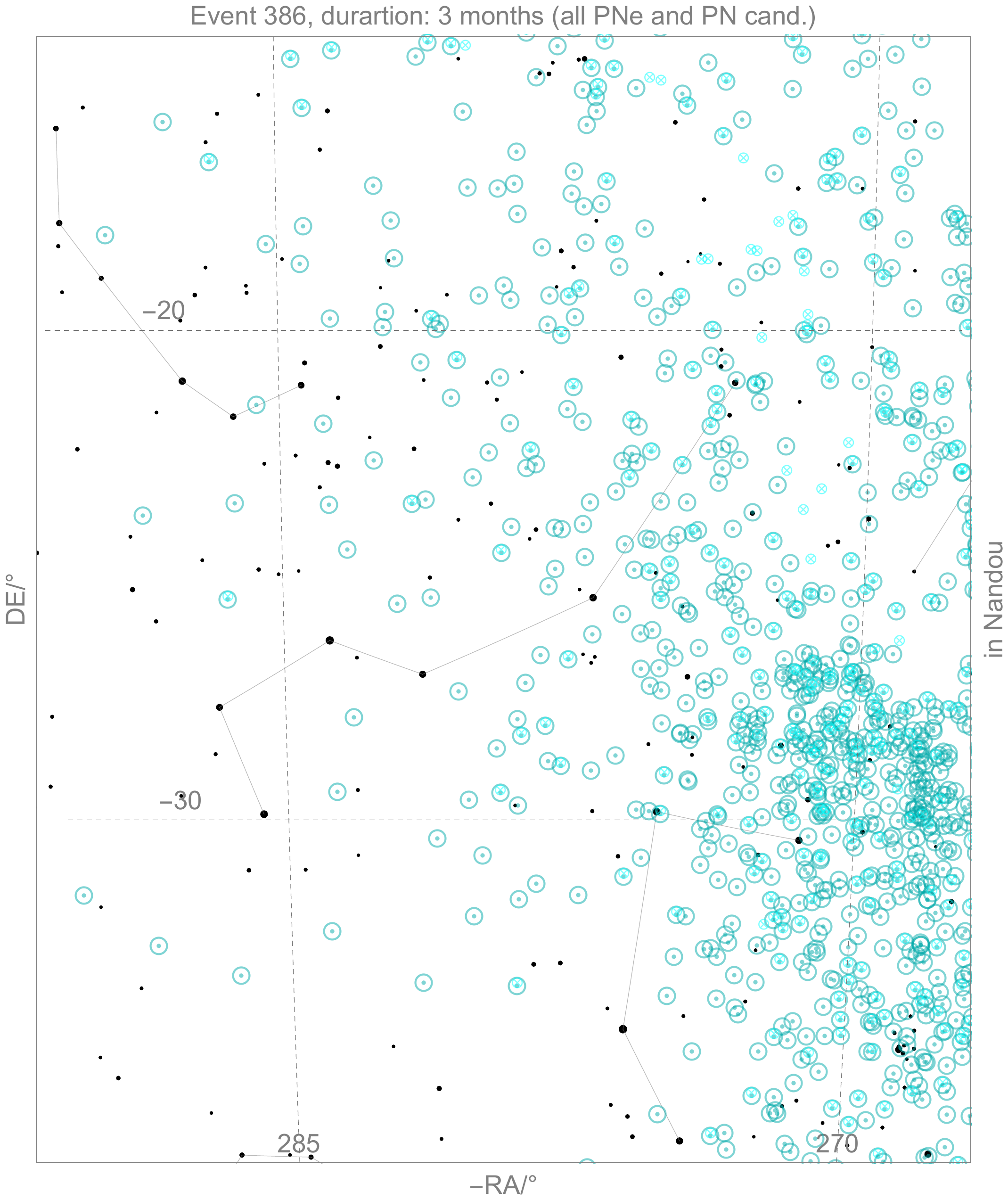} \\
	\includegraphics[width=.98\columnwidth]{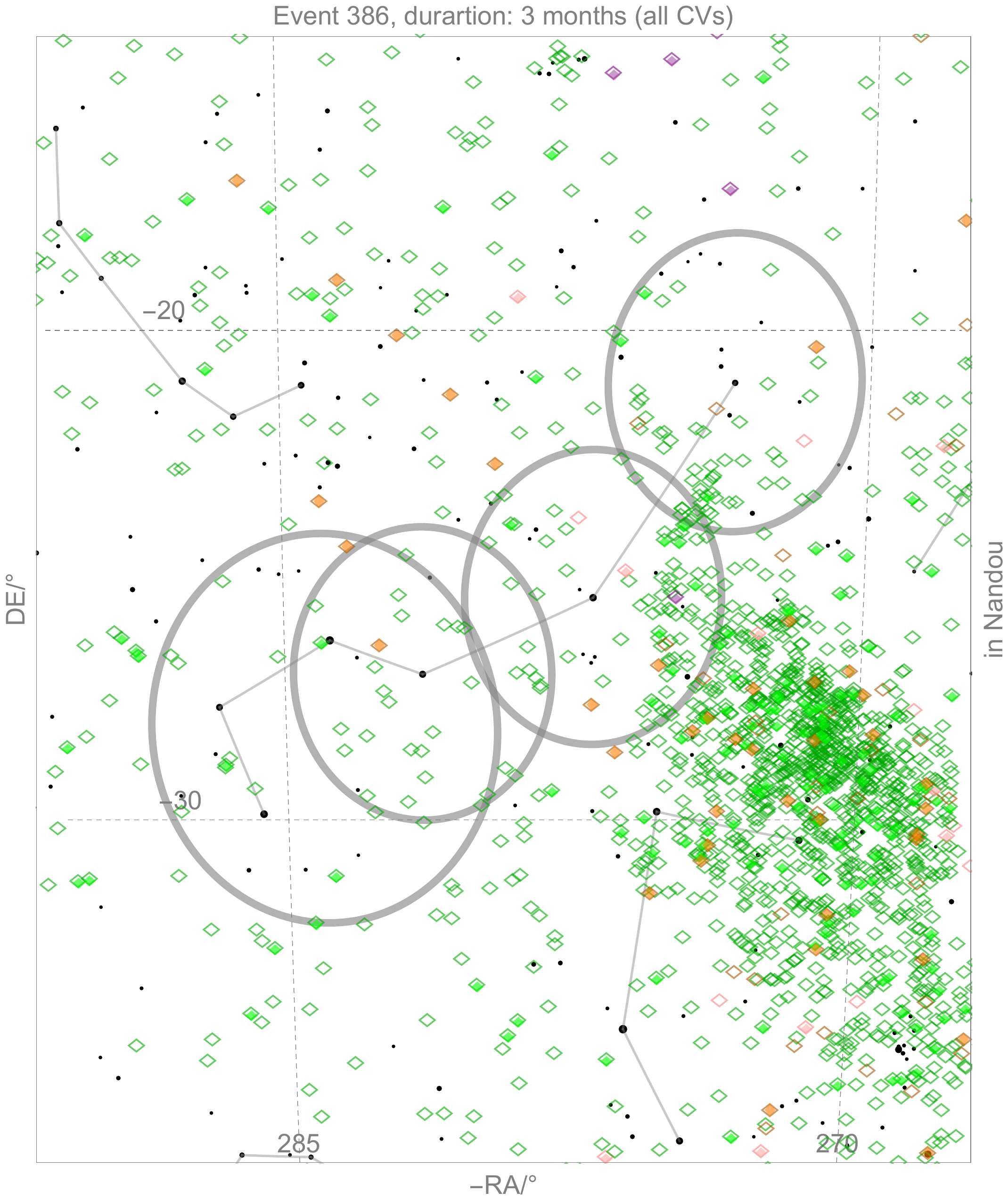}  \hfill
	\includegraphics[width=.98\columnwidth]{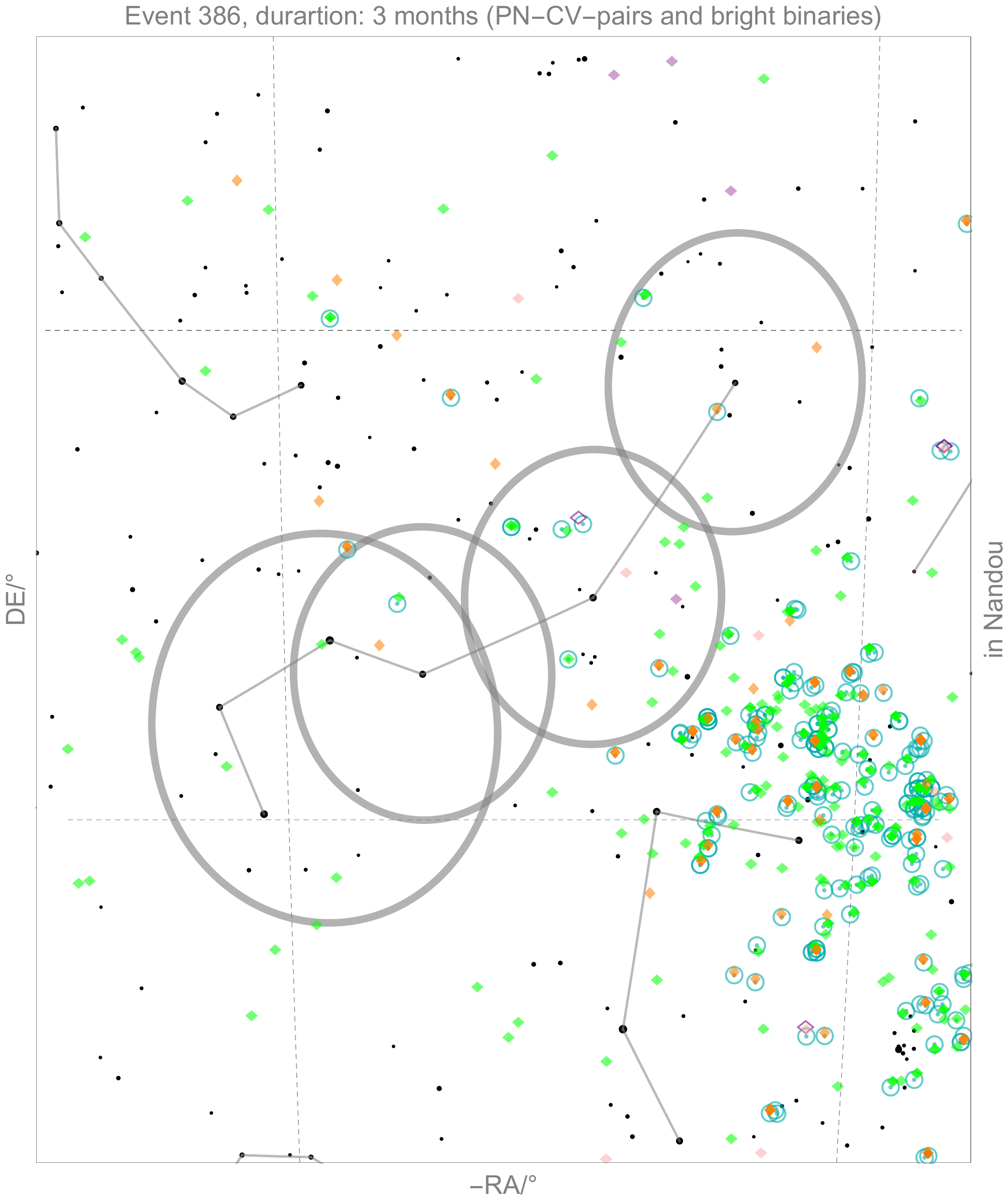} \\
\end{figure*}

 \begin{figure*}
    \caption{Charts for Event 393 (equatorial coordinates, equinox 2000). The four maps display the SNRs `$O$' and PSRs $\ast$ (upper left), the planetary nebulae $\odot,\otimes$ (upper right), all CVs $\diamondsuit$ and the bright ones highlighted (lower left), and the brightest CVs and the planetary nebulae close to them (lower right). The suggested \citep{wang1997} but questionable \citep[p.\,186]{stephGreen} SNR G347.3-00.5 is marked with a red six-pointed star. The grey circle covers the search field in which we suggest to look for a possible counterpart. Their coordinates are displayed in Tab.~\ref{tab:fields}.}
    \label{fig:all393}
    \includegraphics[width=.95\columnwidth]{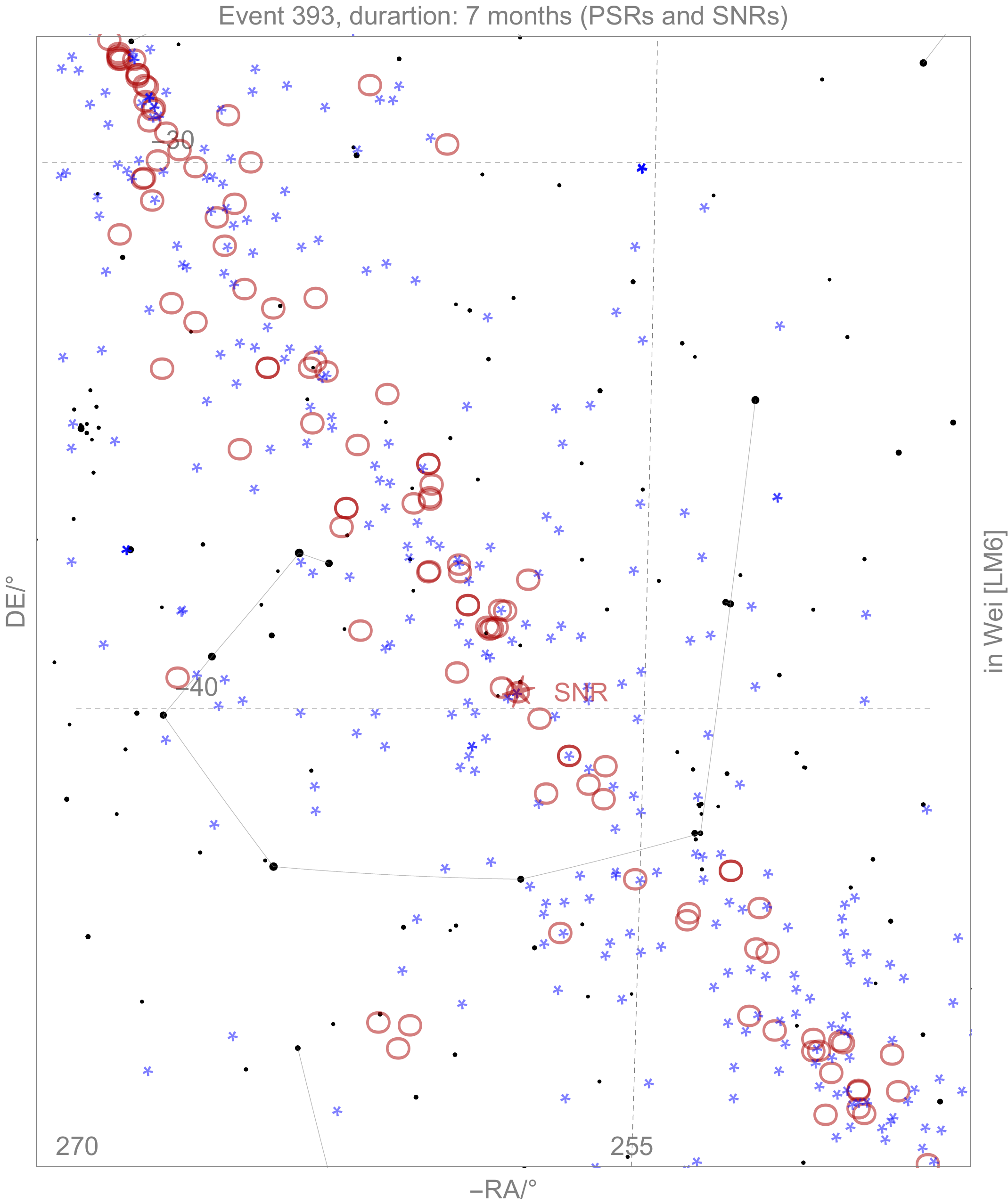}  \hfill
	\includegraphics[width=.95\columnwidth]{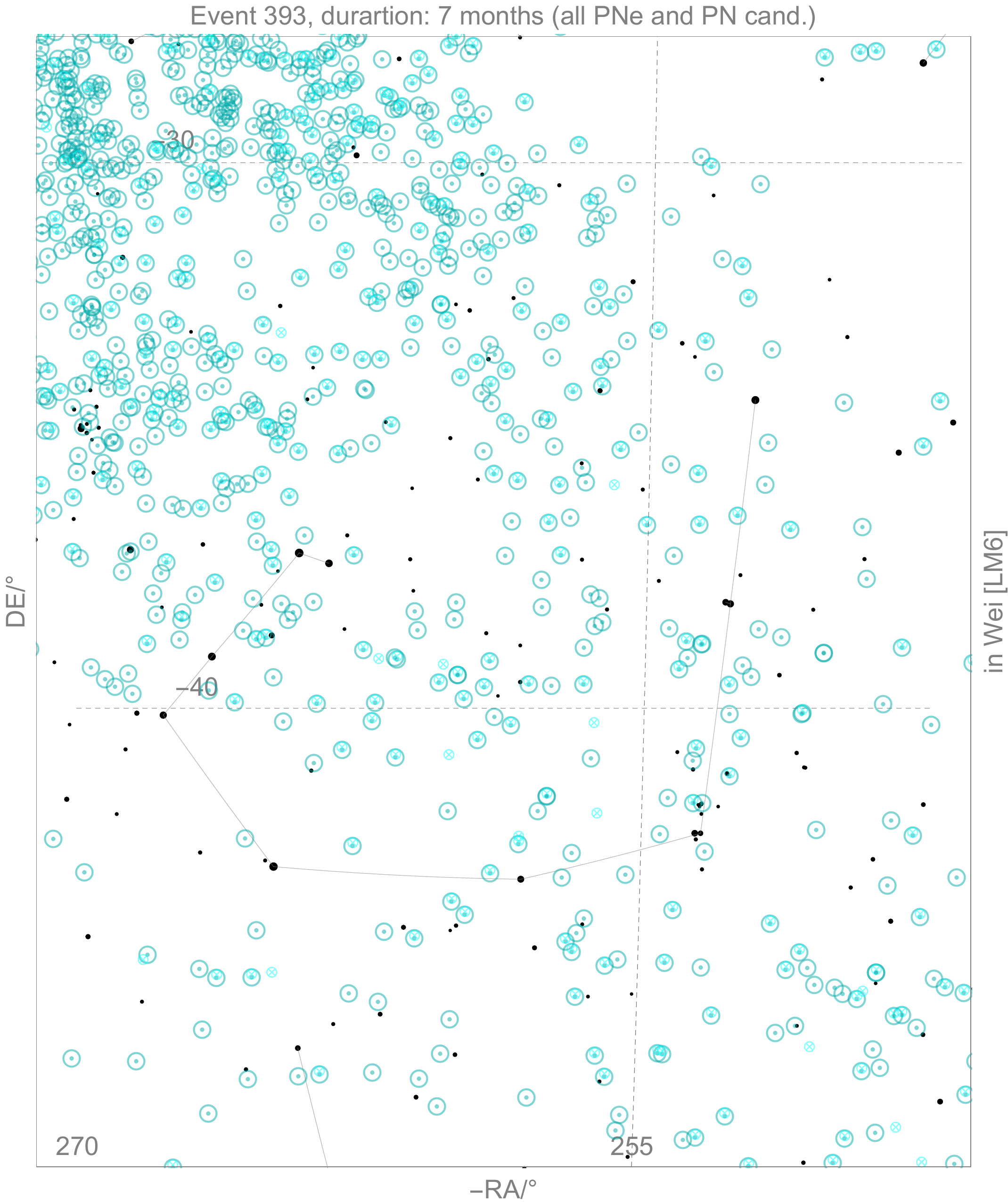} \\
	\includegraphics[width=.95\columnwidth]{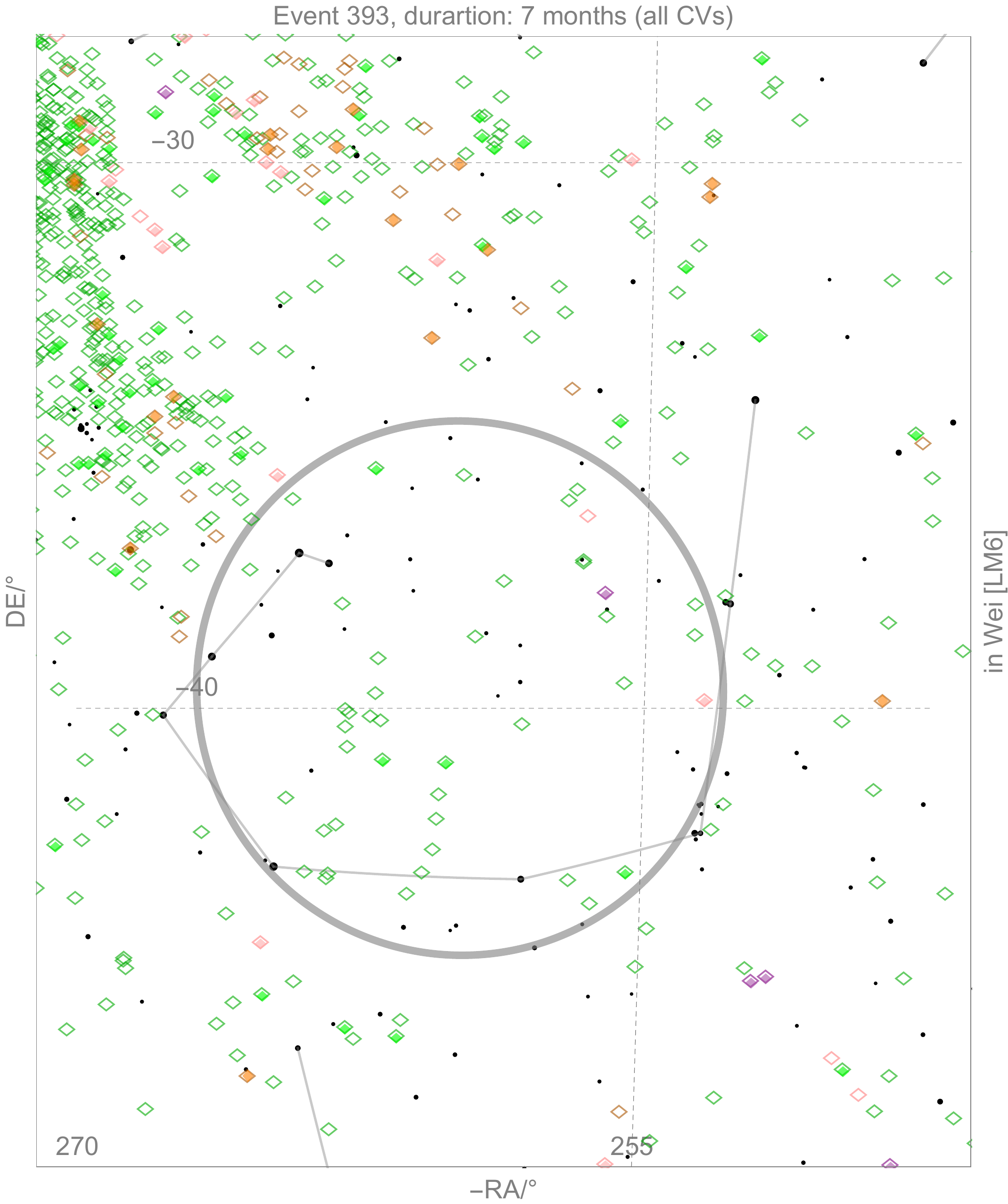} \hfill
	\includegraphics[width=.95\columnwidth]{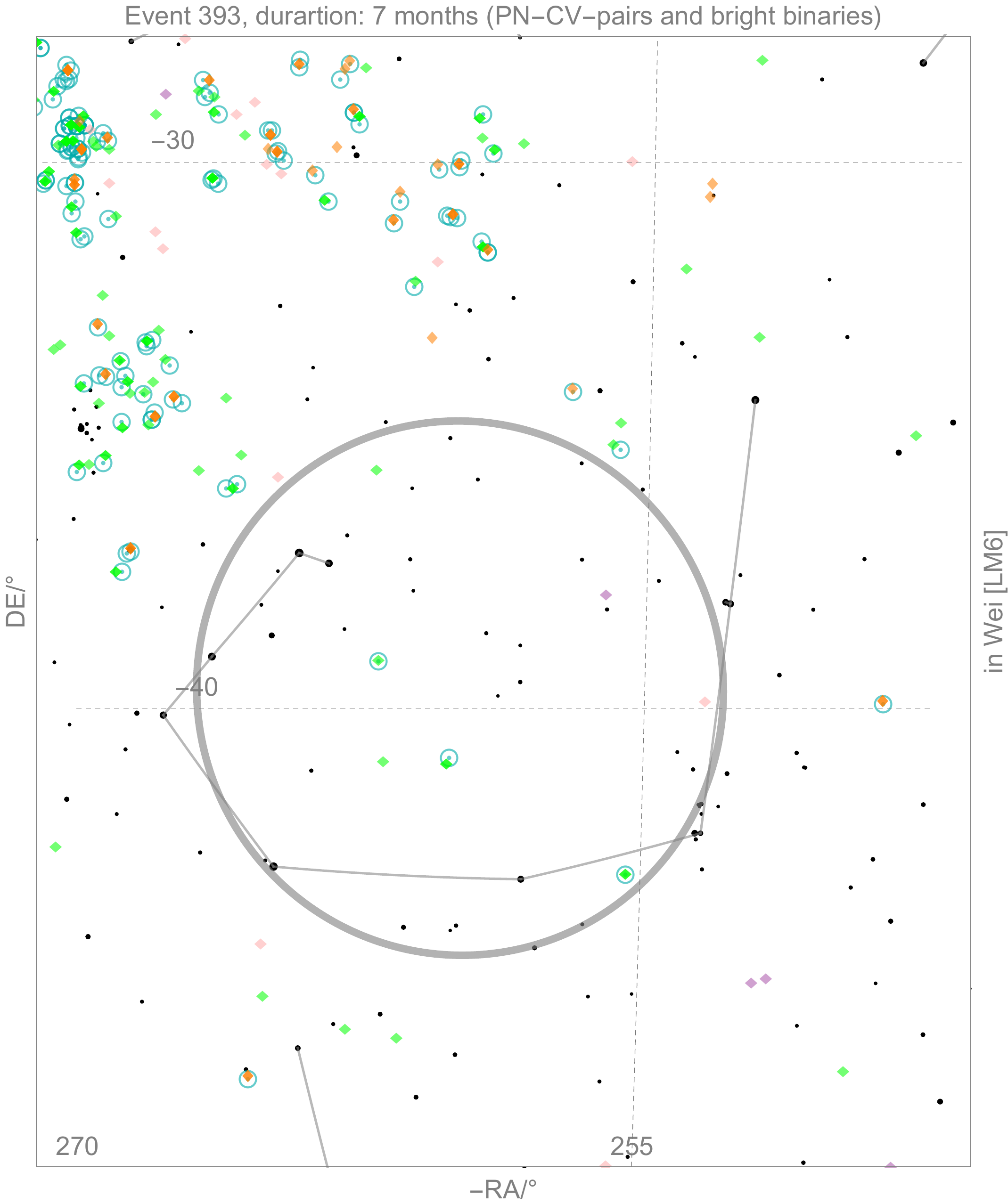} \\
\end{figure*}


\bsp	
\label{lastpage}
\end{document}